\documentstyle[12pt,epsfig]{article}
\def\bge{\begin{equation}}
\def\ene{\end{equation}}
\def\bg{\begin{eqnarray}}
\def\en{\end{eqnarray}}
\def\nn{\nonumber}

\def\del{\partial}

\def\bi{\bibitem}
\begin{document}
\begin{flushright}
ADP-95-45/T194
\end{flushright}
\vspace{0.5cm}
\begin{Large}
\center{THE ROLE OF NUCLEON STRUCTURE IN FINITE NUCLEI}
\end{Large}
\vspace{1.5cm}
\begin{center}
\begin{large}
Pierre~A.~M.~GUICHON~\footnote{pampam@phnx7.saclay.cea.fr} \\
{ \sl DAPHIA/SPhN, CE Saclay, 91191 Gif-sur Yvette, CEDEX, France} \\
\vspace{0.5cm}

Koichi~SAITO~\footnote{ksaito@nucl.phys.tohoku.ac.jp} \\
{\sl Physics Division, Tohoku College of Pharmacy \\
Sendai 981, Japan} \\
\vspace{0.5cm}

Evguenii~RODIONOV~\footnote{erodiono@physics.adelaide.edu.au} and 
Anthony~W.~THOMAS~\footnote{athomas@physics.adelaide.edu.au} \\
{\sl Department of Physics and Mathematical Physics, \\
University of Adelaide, South Australia 5005, Australia}
\end{large}
\end{center}
\vspace{2.0cm}
\newpage
\begin{abstract}
The  quark-meson coupling model, 
based on a mean field description of 
non-overlapping nucleon bags bound by the self-consistent exchange of 
$\sigma$, $\omega$ and $\rho$ mesons, is extended to investigate 
the properties of finite nuclei. Using
the Born-Oppenheimer approximation to describe the 
interacting quark-meson system, we derive the effective 
equation of motion   for 
the nucleon, as well as the self-consistent equations for the 
meson mean fields. 
The model is first applied to nuclear matter, after which we show some 
initial results for finite nuclei.
 
\end{abstract}
%
%
\newpage
\section{Introductory remarks}
\label{sec:Intro}

The nuclear many-body problem has been the object of enormous theoretical
attention for decades. Apart from the non-relativistic
treatments based upon realistic two-body forces~\cite{john,wff}, there
are also studies of three-body effects and higher~\cite{dim}.
The importance of relativity has been recognised in a host of treatments
under the general heading of
Dirac-Brueckner~\cite{cels,nhm,boer}. This approach has
also had considerable success in the treatment of scattering
processes~\cite{rhc}. At the same time, the simplicity of 
Quantum Hadrodynamics (QHD)~\cite{walec,serot} 
has  led to its widespread application, and
to attempts to incorporate the density dependence of the
couplings~\cite{frmu,fuchs} that seems to be required empirically.

One of the fundamental, unanswered questions in this field concerns the
role of sub-nucleonic degrees of freedom (quarks and gluons) in
determining the equation of state. There is little doubt that, at
sufficiently high density (perhaps $5-10\rho_0$, 
with $\rho_0$ the saturation density of symmetric nuclear matter),   
quarks and gluons must
be the correct degrees of freedom and major experimental programs in
relativistic heavy ion physics are either planned or
underway~\cite{ko,expt} to look for this transition. In order to
calculate the properties of neutron stars~\cite{brown} one
needs an equation of state from very low density to many times $\rho_0$
at the centre. A truly consistent theory describing the transition from
meson and baryon degrees of freedom to quarks and gluons might be
expected to incorporate the internal quark and gluon degrees of freedom
of the particles themselves. Our investigation may be viewed as a first
step in this direction. We shall work with the quark-meson coupling (QMC) 
model originally proposed by one of us~\cite{guichon} and since developed
extensively~\cite{st1,hadron}. Related work has been carried out by
a number of groups~\cite{yazaki}. 

Within the QMC model~\cite{guichon} the properties of 
nuclear matter 
are determined by the self-consistent coupling of scalar ($\sigma$) and
vector ($\omega$) fields to the {\it quarks within the nucleons}, 
rather than to the nucleons themselves.
As a result of the scalar coupling the internal structure of the nucleon
is modified with respect to the free case. In particular, the small mass
of the quark means that the lower component of its wave function
responds rapidly to the $\sigma$ field, with a consequent decrease in the
scalar density. As the scalar density is itself the source of the
$\sigma$ field this provides a mechanism for the saturation of nuclear
matter where the quark structure plays a vital role.

In a simple model where nuclear matter was considered as a collection of
static, non-overlapping bags it was shown that a satisfactory
description of the bulk properties of nuclear matter can be 
obtained~\cite{guichon,st1}. 
Of particular interest is the fact that the extra degrees of freedom,
corresponding to the internal structure of the nucleon, result in a
lower value of the incompressibility of nuclear matter than obtained in
approaches based on point-like nucleons -- such as   
QHD~\cite{serot}.  In fact, the prediction is in
agreement with the experimental value once the binding energy and
saturation density are fixed. Improvements to the model, including the
addition of Fermi motion, have not
altered the dominant saturation mechanism.  Furthermore, it is possible to 
give a clear understanding of the relationship between this model and 
QHD~\cite{st1} and to study variations of hadron properties in nuclear 
matter~\cite{hadron}. 
Surprisingly the model seems
to provide a semi-quantitative explanation of the Okamoto-Nolen-Schiffer
anomaly~\cite{ok} when quark mass differences are included~\cite{st3}. 
A further application of the model, including quark mass differences, has
suggested a previously unknown correction to the extraction of the
matrix element, $V_{ud}$, from super-allowed Fermi beta-decay~\cite{wilk}.
Finally the model has been
applied to the case where quark degrees of freedom
are undisputedly involved -- namely the nuclear EMC effect~\cite{st2}. 

Because the model has so far been constructed for infinite nuclear
matter its application has been limited to situations which either
involve bulk properties or where the local density approximation 
has some validity. Our aim here is to overcome this limitation by
extending the model to finite nuclei. In general this is a very
complicated problem and our approach will be essentially classical.
Our starting point will be exactly as for nuclear matter. That is, we
assume that on average the quark bags do not overlap and that the quarks
are coupled locally to average $\sigma$ and $\omega$ fields. The latter
will now vary with position, while remaining time independent -- as they
are mean fields. For deformed or polarised nuclei one should also
consider the space components of the vector field. In order to simplify
the present discussion we restrict ourselves to spherical, spin-saturated
nuclei.  

Our approach to the problem will be within the framework of the 
Born-Oppenheimer approximation. Since the quarks typically move much 
faster than the nucleons we assume that they always have time to adjust 
their motion so that they are in the lowest energy state. 
In order to account for minimal relativistic effects it is convenient 
to work in the instantaneous rest frame (IRF) of the nucleon. 
Implicitly one then knows both the position and the momentum of the nucleon, 
so that the treatment of the motion of the nucleon is classical -- at least, 
as long as the quarks are being considered explicitly. The quantisation of 
the motion of the nucleon is carried out after the quark degrees of freedom 
have been eliminated.

In Sect.~\ref{sec:BO} we show how the 
Born-Oppenheimer approximation can be used to treat the quark degrees of 
freedom in a finite nucleus.  
In Sect.~\ref{sec:quant} we derive the classical equation of motion for a bag 
in the meson fields, including the spin-orbit interaction which is treated in 
first order in the velocity. We then quantize the nucleon motion in a non 
relativistic way. 
The self consistent equations for the meson fields are derived in
Sect.~\ref{sec:meanfield}. 
In Sect.~\ref{sec:self} we summarize the model in 
the form of a self-consistent procedure. The use of this non relativistic 
formulation is postponed to future work. To allow a clear comparison with QHD 
we propose a relativistic formulation in Sect.~\ref{sec:relf}, 
which is then applied to nuclear matter 
in Sect.~\ref{sec:appl}. 
Some initial results for finite nuclei are also presented.  
Sect.~\ref{sec:summ} summarises 
our main results and identifies directions for future work.

\section{The Born-Oppenheimer approximation}
\label{sec:BO}
\setcounter{footnote}{0}

In what follows the coordinates in the rest frame of the nucleus (NRF) 
will be 
denoted without primes: $(t,\vec{r})$. In this frame the nucleon follows 
a classical trajectory, $\vec{R}(t)$. Denoting the instantaneous velocity 
of the nucleon as $\vec{v} = d \vec{R}/d t$, we can define an instantaneous 
rest frame for the nucleon at each time $t$. The coordinates in this IRF 
are $(t',\vec{r}\,'$):
\bg
r_L &=& r_L ' \cosh\xi + t ' \sinh\xi ,  \nonumber  \\
\vec{r}_\perp &=& \vec{r}_\perp\,' ,  \\
t &=& t ' \cosh\xi + r_L ' \sinh\xi , \nn \label{1}
\en 
where $r_L$ and $\vec{r}_\perp$ are the components respectively parallel and  
transverse to the velocity and $\xi$ is the rapidity 
defined by $\tanh\xi = \vert \vec{v}(t) \vert$.

Our assumption that the quarks have time to adjust to the local fields in 
which the nucleon is moving is exact if the fields are constant -- i.e. 
if the motion of the nucleon has no acceleration.
It is, of course, very important to examine the validity of the approximation 
for a typical nuclear environment. For this purpose we take the nucleon 
motion to be non-relativistic. Assume that at time $0$ the nucleon is 
at $\vec{R}_0$. After 
time $t$, assuming $t$ small enough, we have
\bge
\vec{R}(t) = \vec{R}_0 + \vec{v}_0 t + \frac{1}{2} t^2 \vec{\alpha}_0 ,
\label{2}
\ene
where $\vec{\alpha}_0$ ($= \vec{F}/M_N = - \vec{\nabla} V/M_N$) is the 
acceleration and $M_N$ is the free nucleon mass.  We shall take 
the potential to be a typical Woods-Saxon form with depth $V_0 \sim -50$ MeV,
surface thickness $a \sim$ 2 fm and radius $R_A \sim 1.2 A^{1/3}$ fm.
The maximum acceleration occurs at $R = R_A$ and takes the value
\bge
\vec{\alpha}_{max} = V_0 \frac{{\hat R}}{a M_N} .
\label{3}
\ene
Therefore, in the IRF we have
\bge
\vec{R}'(t) \sim \vec{R}_0' + \frac{1}{2} t^2 \frac{V_0}{a M_N} {\hat R}_0 ,
\label{4}
\ene
and, in the worst case, relative to the size of the nucleon itself ($R_B$), 
the departure from a fixed position, is
\bge
\left\vert \frac{R'(t) - R'_0}{R_B} 
\right\vert \sim \frac{1}{2} t^2 \frac{|V_0|}{a M_N R_B} \sim
 \frac {t^2}{80},
\label{5}
\ene
with $t$ in fm.
Thus, as long as the time taken for the quark motion to change is less
than $\sim$ 9 fm, the nucleon position in the IRF can be considered as  
unchanged. 
Since the typical time for an adjustment in the motion of the quarks is 
given by the inverse of the typical excitation energy, which is of order 
0.5 fm, this seems quite safe.

As we have just seen, it is reasonable to describe the internal structure of 
the nucleon in the IRF. In this frame we shall adopt the static spherical
cavity approximation to the MIT bag~\cite{mit} for which the Lagrangian 
density is
\bge
{\cal L}_0 = \bar{q}' ( i \gamma^{\mu} \del_{\mu} -m_q) q' - BV ,
\hspace{0.25truein} \mbox{for } |\vec{u}\,'| \le R_B ,
\label{6}
\ene
with $B$ the bag constant, $R_B$ the radius of the bag, $m_q$ the 
quark mass and $\vec{u}\,'$ the position of the quark from the center of 
the bag (in the IRF): $\vec{u}\,' = \vec{r}\,' - \vec{R}'$. 
We shall denote as $u'$ the 4-vector $(t',\vec{u}\,')$. The field 
$q'(u_0', \vec{u}\,')$ is the quark field in the IRF, which must satisfy the 
boundary condition
\bge
(1 + i \vec{\gamma} \cdot {\hat u}') q' = 0 , 
\hspace{0.25truein} \mbox{at } |\vec{u}\,'| = R_B .
\label{7}
\ene

Next we must incorporate the interaction of the quarks 
with the scalar ($\sigma$) and 
vector ($\omega$) mean fields generated by the other nucleons.
In the nuclear rest frame they are self-consistently generated functions of 
position -- $\sigma(\vec{r})$ and $\omega(\vec{r}$). Using the scalar and 
vector character of these fields, we know that in the IRF their values are
\bg
\sigma_{IRF}(t', \vec{u}\,') &=& \sigma(\vec{r}) , \nonumber \\
\omega_{IRF}(t', \vec{u}\,') &=& \omega(\vec{r}) \cosh\xi , \\
\vec{\omega}_{IRF}(t', \vec{u}\,') &=& - \omega(\vec{r}) \hat{v} \sinh\xi , 
\nn 
\label{8}
\en
and the interaction term is
\bge
{\cal L}_I = g^q_{\sigma} \bar{q}' q' (u') \sigma_{IRF}(u') - g^q_{\omega} 
\bar{q}' \gamma_{\mu} q' (u') \omega^{\mu}_{IRF}(u') ,
\label{9}
\ene
where $g_{\sigma}^q$ and $g_{\omega}^q$ are the 
quark-meson coupling constants for $\sigma$ and $\omega$, respectively. Apart 
from  isospin considerations, the effect of the $\rho$ meson 
can be deduced from the effect of the $\omega$. Thus we postpone its 
introduction to the end of this section.

Since we wish to solve for the structure of the nucleon in the IRF we
need the Hamiltonian and its degrees of freedom in this frame. This
means that the interaction term should be evaluated {\em at equal time}
$t'$ for all points $\vec{u}\,'$ in the bag. Suppose that at time $t'$ the
bag is located at $\vec{R}'$ in the IRF. Then, in the NRF it will be
located at $\vec{R}$ at time $T$ defined by
\bg
R_L &=& R_L' \cosh\xi + t' \sinh\xi , \nonumber \\
\vec{R}_\perp &=& \vec{R}_\perp' , \\
T &=& t' \cosh\xi + R_L' \sinh\xi . \nn 
\label{10}
\en
For an arbitrary point $\vec{r}\,'$ ($ = \vec{u}\,' + \vec{R}'$) in the bag,
at the same time $t'$, we have an analogous relation
\bg
r_L &=& r_L' \cosh\xi + t' \sinh\xi , \nonumber \\
\vec{r}_\perp &=& \vec{r}_\perp\,' , \\
t &=& t' \cosh\xi + r_L' \sinh\xi , \nn 
\label{11}
\en
from which we deduce
\bg
r_L &=& R_L' \cosh\xi + t' \sinh\xi + u_L' \cosh\xi , \nonumber \\
	&=& R_L + u_L' \cosh\xi , \\
\vec{r}_\perp &=& \vec{R}_\perp + \vec{u}_\perp\,' . \nn 
\label{12}
\en
This leads to the following expression for the $\sigma$ field
\bge
\sigma_{IRF}(t', \vec{u}\,') = \sigma(R_L(T) + u_L' \cosh\xi , 
\vec{R}_\perp(T) + \vec{u}_\perp\,') ,
\label{13}
\ene
with a corresponding equation for the $\omega$ field.

The spirit of the Born-Oppenheimer approximation is to solve the
equation of motion for the quarks with the position $\vec{R}(T)$
regarded as a fixed parameter. In order to test the reliability of this
approximation we consider a non-relativistic system and neglect
finite-size effects. That is, we take 
\bge
\sigma_{IRF}(t', \vec{u}\,') \sim \sigma(\vec{R}(T)) .
\label{14}
\ene
As we noted earlier, the typical time scale for a change in the motion of
the quark is $\tau \sim$ 0.5 fm. During this time the relative change of
$\sigma$ due to the motion of the bag is
\bge
\frac{\Delta \sigma}{\sigma} = \vec{v} \cdot {\hat R}
\frac{\sigma'}{\sigma} \tau .
\label{15}
\ene
It is reasonable to assume that $\sigma$ roughly follows the nuclear
density, and as long as this is constant, $\Delta \sigma$ vanishes and the
approximation should be good. The variation of the density occurs mainly
in the surface where it drops to zero from $\rho_0$ (the normal nuclear 
density) over a distance $d\/$ of
about 2 fm. Therefore we can estimate $\vert \sigma' / \sigma \vert$ as
approximately $1/d$ in the region where $\rho$ varies. The factor
$\vec{v} \cdot {\hat R}$ depends on the actual trajectory, but as a rough
estimate we take ${\hat v} \cdot {\hat R} \sim 1/3$. That is, we suppose
that the probability for $\hat{v}$ is isotropic. For the magnitude of
the velocity we take $k_F/M_N$ with $k_F$ = 1.7 fm$^{-1}$ the Fermi
momentum. With these estimates we find 
\bge
\frac{\Delta \sigma}{\sigma} \sim \frac{0.36 \cdot 0.5}{3 \cdot 2} \sim
3\% ,
\label{16}
\ene
which is certainly small enough to justify the use of the
Born-Oppenheimer approximation. Clearly this amounts to neglecting terms
of order $v$ in the argument of $\sigma$ and $\omega$. In order to be
consistent we therefore also neglect terms of order $v^2$ -- that is, we
replace $u_L' \cosh\xi$ by $u_L'$. 
\section{Equation of motion for a bag in the nuclear field}
\label{sec:quant}
\subsection{Leading term in the  Hamiltonian}
\label{susec:leading}
Following the considerations of the previous section, 
in the IRF the interaction
Lagrangian density takes the simple, approximate form:
\bge
{\cal L}_I = g^q_{\sigma} \bar{q}' q' (t', \vec{u}\,') \sigma (\vec{R} + 
\vec{u}\,') - g^q_{\omega} \bar{q}' [\gamma_0 \cosh\xi + \vec{\gamma} \cdot
{\hat v} \sinh\xi] q'(t', \vec{u}\,') \omega (\vec{R} + \vec{u}\,') .
\label{17}
\ene
The corresponding Hamiltonian is~\footnote{Only the quark degrees of
freedom are active. The nucleon position and velocity are parameters as
discussed earlier.} 
\bg
H &=& \int^{R_B}_0 d\vec{u}\,' \, \bar{q}' [- i \vec{\gamma} \cdot 
\vec{\nabla} + m_q - 
g^q_{\sigma} \sigma(\vec{R} + \vec{u}\,')  \nonumber \\
 & & + g^q_{\omega} \omega (\vec{R} +
\vec{u}\,') ( \gamma^0 \cosh\xi + \vec{\gamma} \cdot {\hat v} \sinh\xi ) ]
q'(t', \vec{u}\,') + B V ,
\label{18}
\en
while the momentum is simply
\bge
\vec{P} = \int^{R_B}_0 d\vec{u}\,' \, q'^{\dagger} [- i \vec{\nabla}  ] q' .
\label{19}
\ene
As the $\sigma$ and $\omega$ fields only vary appreciably near the nuclear
surface, where $R \gg |\vec{u}\,'|$ (since $|\vec{u}\,'|$ is bounded by the 
bag radius), it
makes sense to split $H$ into two parts:
\bg
H &=& H_0 + H_1 ,  \label{20} \\
H_0 &=& \int^{R_B}_0 d\vec{u}\,' \, \bar{q}' [ -i \vec{\gamma} \cdot 
\vec{\nabla} + m_q 
- g^q_{\sigma} \sigma(\vec{R}) \nn \\
 & & + g^q_{\omega} \omega(\vec{R}) ( \gamma^0
\cosh\xi + \vec{\gamma} \cdot {\hat v} \sinh\xi ) ] q'(t', \vec{u}\,')
+ BV , \label{21} \\
H_1 &=& \int^{R_B}_0 d\vec{u}\,' \, \bar{q}' [ - g^q_{\sigma} ( \sigma 
(\vec{R} + \vec{u}\,') - \sigma(\vec{R}) )
	\nn \\
 & & + g^q_{\omega} ( \omega (\vec{R} + \vec{u}\,') - \omega
(\vec{R}) ) ( \gamma^0 
\cosh\xi + \vec{\gamma} \cdot {\hat v} \sinh\xi ) ] q'(t', \vec{u}\,') , 
\label{22}
\en
and to consider $H_1$ as a perturbation.

Suppose we denote as $\phi^{\alpha}$ the complete and orthogonal set of 
eigenfunctions defined by
\bg
h \phi^{\alpha}(\vec{u}\,') &\equiv& ( -i \gamma^0 \vec{\gamma} \cdot 
\vec{\nabla} 
 +  m_q^{\star} \gamma^0 ) \phi^{\alpha}(\vec{u}\,') , \nonumber \\
&=& \frac{\Omega_{\alpha}}{R_B} \phi^{\alpha}(\vec{u}\,') , \label{23} \\
(1+i \vec{\gamma}\cdot\hat{u}')\phi^{\alpha}(\vec{u}\,') &=& 0 
 , \hspace{0.25truein} \mbox{at } |\vec{u}\,'| = R_B , \label{24} \\
\int^{R_B}_0 d\vec{u}\,' \, \phi^{\alpha\dagger}\phi^\beta &=& 
\delta^{\alpha\beta} , \label{25}
\en
with $\{\alpha\}$ a collective symbol for the quantum numbers 
and $m_q^{\star}$ a
parameter. Here we recall the expression for the lowest positive energy 
mode, $\phi^{0m}$, with $m$ the spin label: 
\begin{equation}
\phi^{0m}(t', \vec{u}\,') = {\cal N} {j_{0}(xu'/R_B) \choose
i\beta_q {\vec{\sigma}}\cdot \hat{u}' j_{1}(xu'/R_B)}
{\frac{\chi_m}{\sqrt{4\pi}}} , \label{eq:psiq}
\end{equation}
where

\bg
\Omega_0 &=& \sqrt{x^2 + (m_q^{\star}R_B)^2},\ \ \ 
\beta_q = \sqrt{\frac{\Omega_0 - m_q^{\star}R_B}{\Omega_0 + 
m_q^{\star}R_B}} ,\\ 
{\cal N}^{-2} &=& 2R_B^3j^2_0(x)[\Omega_0(\Omega_0 - 1) +
m_q^{\star}R_B/2]/x^2 .  \label{eq:betq}
\en
For this mode, the boundary condition at
the surface amounts to 
\begin{equation}
j_0(x) = \beta_q j_1(x) . \label{eq:eigx}
\end{equation}

We expand the quark  field in the following way
\bge
q'(t', \vec{u}\,') = \sum_\alpha e^{-i\vec k \cdot \vec{u}\,'} 
\phi^\alpha(\vec{u}\,') b_\alpha(t') ,
\label{26}
\ene
with $\vec{k}$ chosen as
\bge
\vec k = g_q^\omega \omega(\vec{R}) \hat{v} \sinh\xi ,
\label{defk}
\ene
in order to guarantee the correct rest frame momentum for a 
particle in a vector field. Substituting into the equation for
$H_0$ we find
\bg
H_0 &=& \sum_{\alpha}\frac{\Omega_\alpha}{R_B} b_\alpha^\dagger b_\alpha
	- \sum_{\alpha\beta}\langle\alpha|(g_\sigma^q \sigma(\vec{R})
-m_q+m_q^{\star})\gamma_0
		|\beta\rangle b_\alpha^\dagger b_\beta \nn \\
     & & + {\hat N}_q g_\omega^q \omega(\vec{R}) \cosh\xi +BV \\
 \vec P &=& \sum_{\alpha\beta}\langle\alpha|-i \vec\nabla|\beta\rangle 
b^\dagger_\alpha b_\beta -  {\hat N}_q \vec k  , 
\label{28}
\en
with the notation
\bge
\langle\alpha|A|\beta\rangle = \int_0^{R_B} d\vec{u}\,' \, 
\phi^{\alpha\dagger}(\vec{u}\,')A\phi^\beta(\vec{u}\,') \hspace{0.25truein} 
\mbox{and} \hspace{0.25truein} 
{\hat N}_q=\sum_\alpha b_\alpha^\dagger b_\alpha . \label{29}
\ene
Choosing $m_q^{\star} = m_q - g_\sigma^q \sigma(\vec{R})$ (in which 
case the frequency $\Omega_\alpha$ and the wave function $\phi^\alpha$ 
become dependent on $\vec{R}$ through $\sigma$) we get for the leading 
part of the energy and momentum operator
\bg
H_0^{IRF} &=& \sum_\alpha \frac{\Omega_\alpha(\vec{R})}{R_B} 
b_\alpha^\dagger b_\alpha  
+ BV + {\hat N}_q  g^q_\omega \omega(\vec{R}) \cosh\xi , 
\label{34} \\
\vec {P}^{IRF} &=& \sum_{\alpha\beta}\langle\alpha|-i \vec\nabla|\beta\rangle 
b^\dagger_\alpha b_\beta - {\hat N}_q  g^q_\omega \omega(\vec{R}) \hat{v} 
\sinh\xi . 
\label{35}
\en
If we quantize the $b_\alpha$ in the usual way, we find that the unperturbed
part of $H$ is diagonalised by states of the form $|N_\alpha,N_\beta, \cdots
\rangle$ with $N_\alpha$ the eigenvalues of the number operator 
$b_\alpha^\dagger b_\alpha$ for the mode $\{\alpha\}$. According our 
working hypothesis, the nucleon should be described by three quarks in the 
lowest mode $(\alpha=0)$
and should remain in that configuration as $\vec{R}$ changes. 
As a consequence, in the expression for the momentum, the contribution of the 
gradient acting on $\phi$ averages to zero by parity and we find that the 
leading terms in the expression for the energy and momentum of the nucleon in 
the IRF are:
\bg
E_0^{IRF} &=& M^{\star}_N(\vec{R})+ 3 g^q_\omega \omega(\vec{R}) \cosh\xi , 
\label{34IRF} \\
\vec {P}^{IRF} &=&  - 3 g^q_\omega \omega(\vec{R}) \hat{v} \sinh\xi , 
\label{35IRF}
\en
with
\bge
M^{\star}_N(\vec{R})=\frac{3\Omega_0(\vec{R})}{R_B}+BV .
\label{35''}
\ene

Since we are going to treat the corrections to leading order in the velocity, 
they will not be affected by the boost back to the NRF. 
Therefore  we can  apply the Lorentz transformation to 
Eqs.(\ref{34IRF}) and (\ref{35IRF}) to get the leading terms in 
the energy and 
momentum in the NRF. Since there is no possibility of confusion now, 
we write the NRF variables without the NRF index. The result is
\bg
E_0 &=& M^{\star}_N(\vec{R})\cosh\xi+ 3 g^q_\omega \omega(\vec{R}) , 
\label{34NRF} \\
\vec {P} &=& M^{\star}_N(\vec{R})\hat{v} \sinh\xi  , 
\label{35NRF}
\en
which implies
\bge
E_0=\sqrt{M^{\star 2}_N(\vec{R})+\vec {P}^2}+3 g^q_\omega \omega(\vec{R}) .
\label{36NRF}
\ene

At this point we recall that the effective mass of the nucleon, 
$M_N^{\star}(\vec{R})$, 
defined by Eq.(\ref{35''}) does not take into account the fact that the 
center of mass of the quarks does not coincide with the center of the bag. 
By requiring that all of the quarks remain in the same orbit one forces 
this to be realized in expectation value. However, one knows that the virtual 
fluctuations to higher orbits would decrease the energy. This c.m. 
correction is studied in detail in the Appendix, where it is 
shown that it is only very weakly dependent on 
the external field strength for the 
densities of interest. For the zero point energy due to the fluctuations 
of the gluon field, we assume that it is the same as in free space. 
Thus we parametrize the sum of the c.m. and gluon fluctuation corrections 
in the familiar form, $-z_0/R_B$, where $z_0$ is {\it independent} of the 
density. 
Then the effective mass of the nucleon in the nucleus takes the form
\bge
M_N^{\star}(\vec{R}) = \frac{3\Omega_0(\vec{R})-z_0}{R_B}  + BV , \label{42'}
\ene
and we assume that the equilibrium condition is  
\bge
\frac{d M_N^{\star}(\vec{R})}{d R_B} = 0, \label{42''}
\ene
which is the usual non-linear boundary condition. 
This is again justified by the Born-Oppenheimer 
approximation, according to which  the internal structure of the nucleon has 
enough time to adjust to the varying external field so as to 
stay in its ground state.

The parameters $B$ and $z_0$ are  fixed by the free nucleon mass ($M_N$ = 
939 MeV) using  Eqs.(\ref{42'}) and (\ref{42''}) applied to the free case. 
In the following we keep the free bag radius, $R^0_B$, as a free parameter. 
The results are shown in Table~\ref{b,z}. 
\begin{table}[hbtp]
\begin{center}
\caption{$B^{1/4}$(MeV) and $z_0$ for some bag radii using $m_q$ = 5 MeV.}
\label{b,z}
\begin{tabular}[t]{cccc}
\hline
$R^0_B$(fm) & 0.6 & 0.8 & 1.0 \\
\hline
$B^{1/4}$ & 210.9 & 170.0 & 143.8 \\
$z_0$     & 4.003 & 3.295 & 2.587 \\
\hline
\end{tabular}
\end{center}
\end{table}

\subsection{Corrections due to $H_1$ and Thomas precession}
\label{correction}
We estimate the effect of $H_1$ in perturbation theory by 
expanding $\sigma(\vec{R} + \vec{u}\,')$ and $\omega(\vec{R} + \vec{u}\,')$ 
in powers of $\vec{u}\,'$ and computing the effect to first order. 
To this
order several terms give zero because of parity and one is left with
\bge
\langle(0)^3|H_1|(0)^3\rangle = g_\omega^q \sum_{\alpha\beta} 
\langle(0)^3|b_\alpha^\dagger b_\beta|(0)^3\rangle \langle\alpha| 
\gamma^0{\vec\gamma}\cdot\hat{v} \vec{u}\,' \sinh\xi |\beta\rangle
\cdot\vec\nabla_R \omega(\vec{R}) .
\label{37}
\ene
Here we need to be more precise about the meaning of the labels $\alpha, 
\beta$.
Suppose we set $\{\alpha\}=\{0,m_\alpha\}$ 
with $m_\alpha$ the spin projection of the 
quark in mode $\{0\}$,
then we find
\bge
\langle(0) m_\alpha| \gamma^0\vec\gamma\cdot\hat{v} \vec{u}\,' \sinh\xi|(0)
 m_\beta\rangle = - I(\vec{R}) \langle m_\alpha|\vec \frac{\sigma}{2} 
|m_\beta\rangle 
\times {\hat v} \sinh\xi ,
\label{38}
\ene
with 
\bg
I(\vec{R}) &=& \frac{4}{3}{\cal N}^2 \int_0^{R_B} du' \, u'^3 j_0(xu'/R_B) 
\beta_q j_1(xu'/R_B), \label{Idef} \\
  &=& \frac{R_B}{3} \left( \frac{4\Omega_0 + 2m_q^{\star}R_B 
-3}{2\Omega_0(\Omega_0-1)+m_q^{\star}R_B} \right) . \label{Idef2}
\en

The integral, $I(\vec{R})$, depends on $\vec{R}$ through the implicit 
dependence of $R_B$ and $x$ on the local scalar field. 
Its value in the free case, $I_0$, can be related to the nucleon isoscalar 
magnetic 
moment: $I_0=3 \mu_s/M_N$ with $\mu_s=\mu_p+\mu_n$ and $\mu_p=2.79$, 
$\mu_n=-1.91$ the experimental values. 
By combining Eqs.(\ref{37}), (\ref{38}) and (\ref{35NRF}) we then find
\bg
\langle(0)^3|H_1|(0)^3\rangle &=& \mu_s \frac{I(\vec{R})}{I_0}
\frac{1}{M_N R} \left( \frac{d}{dR} 3 g_\omega^q\omega(\vec{R}) \right) 
\vec{S} \cdot{\vec R}\times{\hat v} \sinh\xi\\
&=& \mu_s \frac{I(\vec{R})M^{\star}_N(\vec{R})}{I_0M_N}
\frac{1}{M^{\star 2}_N(\vec{R}) R} \left( \frac{d}{dR} 
3 g_\omega^q\omega(\vec{R}) \right) \vec{S} \cdot\vec{L},
\label{39}
\en
with $\vec{S}$ the nucleon spin operator and $\vec{L}$ its angular momentum.

This spin-orbit interaction is nothing but the interaction between the 
magnetic moment of the nucleon with the ``magnetic" field of
the $\omega$ seen from the rest frame of the nucleon.
This  induces a rotation of the spin as a function of time. However, even if 
$\mu_s$ were equal to zero, the spin would nevertheless rotate because of  
Thomas precession, which is a relativistic effect {\it independent} of the 
structure. It can be understood as follows. 

Let us assume that at time $t$, the spin vector is $\vec{S}(t)$ 
in the  IRF($t$). Then we expect that, at time $t+dt$ the spin 
has the same direction if it is viewed from the frame  
obtained by boosting the IRF($t$) by $d\vec{v}$ so as to get the right 
velocity $\vec{v}(t+dt)$. That is, the spin looks at rest in the frame 
obtained 
by first boosting the NRF to $\vec{v}(t)$ and then boosting by $d\vec{v}$. 
This product of Lorentz transformation amounts to a boost to $\vec{v}(t+dt)$ 
times a rotation. So, viewed from the IRF($t+dt$), the spin appears to 
rotate. In order that our Hamiltonian be correct it should contain a  
piece $H_{prec}$ which produces this rotation through the Hamilton 
equations of motion. A detailed derivation can be found in 
Refs.\cite{jackson,goldstein} and the result is
\bge
H_{prec}=-\frac{1}{2}\vec{v}\times\frac{d\vec{v}}{dt} \cdot \vec{S}.
\label{39a}
\ene
The acceleration is obtained from the Hamilton equations of motion applied to 
the leading order  Hamiltonian, Eq.(\ref{36NRF}). To lowest order in the 
velocity one finds 
\bge
\frac{d\vec{v}}{dt}=-\frac{1}{M_N^{\star}(\vec{R})} 
\vec{\nabla}[M_N^{\star}(\vec{R})+3g_\omega^q\omega(\vec{R})].
\label{39b}
\ene
If we put this result into Eq.(\ref{39a}) and add the result  to 
Eq.(\ref{39}),  we get the total spin orbit interaction (to first order 
in the velocity)
\bge
H_{prec} + H_1 = V_{s.o.}({\vec R}) \vec{S} \cdot\vec{L}, 
\label{39ca}
\ene
where
\bge
V_{s.o.}({\vec R}) = -\frac{1}{2M^{\star 2}_N(\vec{R}) R} 
\left[ \left( \frac{d}{dR} M^{\star}_N(\vec{R}) \right) 
+(1-2\mu_s\eta_s(\vec{R})) \left( \frac{d}{dR} 3g_\omega^q\omega(\vec{R}) 
\right) \right], 
\label{39cb}
\ene
and 
\bge
\eta_s(\vec{R}) = \frac{I(\vec{R})M_N^{\star}(\vec{R})}{I_0 M_N}. 
\label{39cc}
\ene
\subsection{Total Hamiltonian for a bag in the meson mean fields}
To complete the derivation we now introduce the effect of the neutral 
$\rho$ meson. The interaction term that we must add to ${\cal L}_I$ 
(see Eq.(\ref{9})) is
\bge
{\cal L}_I^{\rho} =  - g^q_{\rho} \bar{q}'
\gamma_{\mu}\frac{\tau_\alpha}{2} q' (u') \rho^\mu_{\alpha,IRF}(u') ,
\label{46'}
\ene
where $\rho^\mu_{\alpha,IRF}$ is  the $\rho$ meson field with isospin 
component $\alpha$ and $\tau^\alpha$ are the Pauli matrices acting on the 
quarks. In the mean field approximation only $\alpha =3$ contributes. 
If we denote by $b(\vec{R})$ the mean value of the time component of 
the field in the NRF, we can transpose our  results for the $\omega$ field. 
The only difference comes from trivial isospin factors which amount 
to the substitutions
\bge
3g^q_\omega\to g^q_\rho\frac{\tau^N_3}{2},\ \ \mu_s\to \mu_v=\mu_p-\mu_n,
\label{46''}
\ene
where $\tau^N_3/2$ (with eigenvalues $\pm 1/2$) 
is the nucleon isospin operator. 

This leads to our final result for the NRF energy-momentum of the 
nucleon moving in the mean fields, 
$\sigma(\vec{R}), \omega(\vec{R})$ and $b(\vec{R})$: 
\bg
E &=& M_N^{\star}(\vec{R}) \cosh\xi + V(\vec{R}), \label{46.3} \\
\vec P &=& M_N^{\star}(\vec{R}) \hat{v} \sinh\xi,\label{46.4}
\en
with
\bg
V(\vec{R}) &=& V_c(\vec{R})+V_{s.o.}(\vec{R})\vec{S} \cdot\vec{L},
\label{VR} \\ 
V_c(\vec{R}) &=& 3 g_\omega^q \omega(\vec{R}) +g_\rho^q \frac{\tau_3^N}{2} 
b(\vec{R}), \label{46.5}\\
V_{s.o.}({\vec R}) &=& -\frac{1}{2M^{\star 2}_N(\vec{R}) R}[\Delta_\sigma
  +(1-2\mu_s\eta_s(\vec{R}))\Delta_\omega
+ (1-2\mu_v\eta_v(\vec{R}))\frac{\tau^N_3}{2}\Delta_\rho ], 
\label{46.06} \\
\Delta_\sigma &=& \frac{d}{dR} M^{\star}_N(\vec{R}),\ 
\Delta_\omega = \frac{d}{dR} 3g_\omega^q\omega(\vec{R}),\  
\Delta_\rho =\frac{d}{dR} g_\rho^q b(\vec{R}) .
\label{46.6}
\en
Up to terms of higher order in the velocity, this  result for the spin-orbit 
interaction agrees  with that  of Achtzehnter and Wilets~\cite{larry} 
who used an approach based on the non-topological soliton model. As 
pointed out in \cite{larry}, for a point-like Dirac particle one 
has $\mu_s=1$ while the physical value is $\mu_s=.88$. Thus, in so far as the 
omega contribution to the spin-orbit force is concerned, the point-like 
result is almost correct. This is clearly  not the case for the rho 
contribution since we still have $\mu_v=1$ for the point-like 
particle while experimentally $\mu_v=4.7$.  
\subsection{Quantization of the nuclear Hamiltonian}
Until now the motion of the nucleon has been considered as classical,
but we must now
quantize the model.  We do that here in the non-relativistic framework.
That is, we consider a theory where the particle number is conserved and 
we keep only terms up to those quadratic in the velocity. 
(Here we drop the spin dependent correction since it already involves
the velocity. It is reinserted at the end.)

The simplest way to proceed is to realize that the equations of motion
\bg
E &=& M_N^{\star}(\vec{R}) \cosh\xi + V_c(\vec{R}), \\
\vec P &=& M_N^{\star}(\vec{R}) \hat{v} \sinh\xi,
\en
can be derived from the following Lagrangian  
\bge
L(\vec{R},\vec{v})=- M_N^{\star}(\vec{R})\sqrt{1-v^2}-V_c(\vec{R}).
\label{nuc1}
\ene
Thus the expansion is simply
\bge
L_{nr}(\vec{R},\vec{v})=\frac{1}{2} M_N^{\star}(\vec{R})v^2- 
M_N^{\star}(\vec{R})-V_c(\vec{R})+ {\cal O}(v^4)
\label{nuc2}
\ene
Neglecting the ${\cal O}(v^4)$ terms we then go back to the (still classical) 
Hamilton variables and find (note that the momentum is not the same as before 
the expansion)
\bge
H_{nr}(\vec{R},\vec{P})=\frac{\vec{P}^2}{2 M_N^{\star}(\vec{R})} + 
M_N^{\star}(\vec{R})+V_c(\vec{R}).
\label{nuc3}
\ene
If we quantize by the substitution $\vec{P}\to -i\vec{\nabla}_R$ we meet the 
problem of ordering ambiguities since now $M_N^{\star}(\vec{R})$ 
and $\vec{P}$ no longer commute. The classical kinetic energy term 
$\vec{P}^2/2 M_N^{\star}(\vec{R}) $  allows 3 quantum orderings 
\bge
T_1=\vec{P} \cdot {\hat A} \vec{P},\ \ T_2={\hat A} \vec{P}^2,\ \ 
T_3=\vec{P}^2 {\hat A} ,
\label{nuc4}
\ene
where ${\hat A}=1/(2 M_N^{\star}(\vec{R}))$. However,
since we want the Hamiltonian to be hermitian the only possible 
combination is 
\bge
T=z T_1 +\frac{1-z}{2}(T_2+T_3),
\label{nuc5}
\ene
with $z$ an arbitrary real number. Using the commutation relations one can 
write
\bge
T_2+T_3=2 T_1-\frac{1-z}{2} (\nabla^2_R {\hat A}),
\label{nuc6}
\ene
so that the kinetic energy operator 
\bge
T=T_1-\frac{1-z}{2}(\nabla^2_R {\hat A}),
\label{nuc7}
\ene
is only  ambiguous through the term containing the second derivative of 
${\hat A}=1/(2 M_N^{\star}(\vec{R}))$. 
Since we have  expanded  the meson fields only to first order in the 
derivative to get the classical Hamiltonian it is consistent to neglect 
such terms. So our quantum Hamiltonian takes the form
\bge
H_{nr}(\vec{R},\vec{P})={\vec{P} \cdot \frac{1}{2 M_N^{\star}(\vec{R})}}
\vec{P}+M_N^{\star}(\vec{R})+V(\vec{R}) ,
\label{nuc8}
\ene
where we have reintroduced the spin-orbit interaction in the potential 
$V(\vec{R})$. The nuclear 
quantum Hamiltonian appropriate to a mean field calculation is then
\bge
H_{nr}=\sum_{i=1,A}H_{nr}(\vec{R}_i,\vec{P}_i),\ \ \ 
\vec{P}_i=-i\vec{\nabla}_i.
\label{nuc9}
\ene
The problem is manifestly self consistent because the meson mean fields, upon 
which $H_{nr}$ depends through $M_N^{\star}$ and $V$, 
themselves depend on the eigenstates of $H_{nr}$.  

\section{Equations for the meson fields}
\label{sec:meanfield}

The equation of motion for the meson-field operators 
($\hat{\sigma},\ \hat{\omega}^\nu ,\ \hat{\rho}^{\nu ,\alpha}$) are
\bg
\del_\mu \del^\mu \hat{\sigma} + m_\sigma^2 \hat{\sigma} &=&g_\sigma^q 
\overline{q}q, \label{mf1}\\
\del_\mu \del^\mu \hat{\omega}^\nu + m_\omega^2 \hat{\omega}^\nu &=& 
g_\omega^q \overline{q}\gamma^\nu q ,\label{mf1'}\\
\del_\mu \del^\mu \hat{\rho}^{\nu,\alpha} + m_\rho^2 \hat{\rho}^{\nu,\alpha} 
&=&g_\rho^q \overline{q}\gamma^\nu \frac{\tau^\alpha}{2}q.
\label{mf1''}\en
The mean fields are defined as the expectation values in the ground state of 
the nucleus, $|A \rangle$:
\bg
\langle A|\hat{\sigma}(t, \vec{r})|A \rangle &=&\sigma(\vec{r}),\label{mf2}\\
\langle A|\hat{\omega}^\nu (t, \vec{r})|A \rangle &=&
\delta(\nu,0)\omega(\vec{r}), \label{mf2'}\\
\langle A|\hat{\rho}^{\nu,\alpha} (t, \vec{r})|A \rangle &=&\delta(\nu,0)
\delta(\alpha,3)b(\vec{r}).
\label{mf2''}\en
The equations which determine them are the expectation values of 
Eqs.(\ref{mf1}), (\ref{mf1'}) and (\ref{mf1''}). First we need the 
expectation values of the sources
\bge
\langle A|\overline{q}q (t, \vec{r})|A \rangle, \langle 
A|\overline{q}\gamma^\nu q (t, \vec{r})|A \rangle \mbox{ and } \langle 
A|\overline{q}\gamma^\nu \frac{\tau^\alpha}{2} q (t, \vec{r})|A \rangle 
.\label{mf3}
\ene
As before, we shall simplify the presentation by not treating
the $\rho$ meson explicitly until the end. 
In the mean field approximation the sources are the sums of the sources
created by  each nucleon -- the latter  interacting with the meson fields. 
Thus we write 
\bg
\overline{q}q (t, \vec{r})&=&\sum_{i=1,A} \langle \overline{q}q (t, \vec{r}) 
\rangle_i, \label{mf4}\\
\overline{q}\gamma^\nu q (t, \vec{r})&=&\sum_{i=1,A} \langle 
\overline{q}\gamma^\nu q (t, \vec{r}) \rangle_i, \label{mf4'}
\en
where $\langle \cdots \rangle_i$ denotes the matrix element in the 
nucleon $i$ located 
at $\vec{R}_i$ at time $t$. According to the Born-Oppenheimer approximation, 
the nucleon structure is described, in its own IRF, by 3 quarks in the 
lowest mode. Therefore, in the IRF of the nucleon $i$, we have
\bg
\langle \overline{q}'q'(t', \vec{r}\,') \rangle_i&=& 3 \sum_m 
\overline{\phi}^{0,m}_i(\vec{u}\,')\phi^{0,m}_i(\vec{u}\,')= 3 
s_i(\vec{u}\,'), \label{mf5}\\
\langle \overline{q}'\gamma^\nu q'(t', \vec{r}\,') \rangle_i &=& 3 
\delta(\nu,0) \sum_m 
\phi^{\dagger 0,m}_i(\vec{u}\,')\phi^{0,m}_i(\vec{u}\,')= 3 \delta(\nu,0) 
w_i(\vec{u}\,'),
\label{mf5'}
\en
where the space components of the vector current gives zero  because of
parity. At the common time $t$ in the NRF we thus have
\bg
R_{i,L}'&=& R_{i,L} \, \cosh\xi_i 
- t \, \sinh\xi_i ,\ \ \vec{R}_{i,\perp}'=\vec{R}_{i,\perp} ,\label{mf6}\\
r_L'&=& r_L \, \cosh\xi_i - t \, \sinh\xi_i ,\ \ 
\vec{r}_{\perp}\,'=\vec{r}_{\perp}. \label{mf6'}
\en
Therefore
\bge
u_{i,L}'= (r_L-R_{i,L}) \cosh\xi_i , \hspace{0.25truein} 
\vec{u}_{i,\perp}\,'= 
\vec{r}_\perp -\vec{R}_{i,\perp} ,\label{mf7}
\ene
and from the Lorentz transformation properties of the fields we get
\bg
\langle \overline{q}q (t, \vec{r}) \rangle_i &=& 3 s_i((r_L-R_{i,L}) 
\cosh\xi_i, \vec{r}_\perp -\vec{R}_{i,\perp}) ,\label{mf8}\\
\langle \overline{q}\gamma^0 q (t, \vec{r}) \rangle_i &=& 3 w_i((r_L-R_{i,L}) 
\cosh\xi_i, \vec{r}_\perp -\vec{R}_{i,\perp}) \cosh\xi_i ,\label{mf8'}\\
\langle \overline{q}\vec{\gamma} q (t, \vec{r}) \rangle_i &=& 3 
w_i((r_L-R_{i,L})\cosh\xi_i, \vec{r}_\perp -\vec{R}_{i,\perp})\hat{v}_i 
\sinh\xi_i . 
\label{mf8''}
\en
These equations can be re-written in the form
\bg
\langle \overline{q}q (t, \vec{r}) \rangle_i &=& \frac{3}{(2\pi)^3} 
(\cosh\xi_i)^{-1} \int d\vec{k} \, e^{i\vec{k}\cdot(\vec{r}-\vec{R}_i)}\ 
S(\vec{k}, \vec{R}_i), \label{mf9}\\
\langle \overline{q}\gamma^0 q (t, \vec{r}) \rangle_i &=& \frac{3}{(2\pi)^3} 
\int d\vec{k} \, e^{i\vec{k}\cdot(\vec{r}-\vec{R}_i)}\ W(\vec{k}, \vec{R}_i), 
\label{mf9'}\\
\langle \overline{q}\vec{\gamma} q (t, \vec{r}) \rangle_i &=& 
\frac{3}{(2\pi)^3} \vec{v}_i \int d\vec{k} \, 
e^{i\vec{k}\cdot(\vec{r}-\vec{R}_i)}\ W(\vec{k}, \vec{R}_i),
\label{mf9''}
\en
with 
\bg
S(\vec{k},\vec{R}_i)=\int d\vec{u} \, e^{-i(\vec{k}_\perp \cdot\vec{u}_\perp 
+k_Lu_L/\cosh\xi_i)}\ s_i(\vec{u}),\label{mf10}\\
W(\vec{k},\vec{R}_i)=\int d\vec{u} \, e^{-i(\vec{k}_\perp \cdot\vec{u}_\perp 
+k_Lu_L/\cosh\xi_i)}\ w_i(\vec{u}).
\label{mf10'}
\en
Finally, the mean field expressions for the meson sources take the form
\bg
\langle A|\overline{q}q (t, \vec{r})|A \rangle &=& \frac{3}{(2\pi)^3} \int 
d\vec{k} \, e^{i\vec{k}\cdot\vec{r}} \langle A|\sum_i (\cosh\xi_i 
)^{-1}e^{-i\vec{k}\cdot\vec{R}_i}\ S(\vec{k}, \vec{R}_i)|A \rangle, 
\label{mf11}\\
\langle A|\overline{q}\gamma^0 q (t, \vec{r})|A \rangle &=& 
\frac{3}{(2\pi)^3} 
\int d\vec{k} \, e^{i\vec{k}\cdot\vec{r}} \langle A|\sum_i 
e^{-i\vec{k}\cdot\vec{R}_i}\ W(\vec{k}, \vec{R}_i)|A \rangle ,\label{mf11'}\\
\langle A|\overline{q}\vec{\gamma} q (t, \vec{r})|A \rangle &=& 0 ,
\label{mf11''}
\en
where the last equation follows from the fact that the velocity vector 
averages to zero.

To simplify further, we remark that a matrix element of the form
\bge
\langle A|\sum_ie^{-i\vec{k}\cdot\vec{R}_i}\cdots|A \rangle  \label{mf12}
\ene
is negligible unless $k$ is less than, or of the order of, the reciprocal of
the {\it nuclear} radius. But in Eqs.(\ref{mf10}) and (\ref{mf10'}) 
$\vec k$ is 
multiplied by $\vec u$ which is bounded by the {\it nucleon} radius. 
Hence, if we restrict the application of the model to large enough nuclei, 
we can neglect  the argument of the exponential in Eqs.(\ref{mf10}) and 
(\ref{mf10'}). The evaluation of the correction to this approximation  
will be postponed to a future work.

We now define the  scalar, baryonic and isospin densities of the
nucleons in the nucleus by
\bg
\rho_s(\vec{r})&=& \langle A|\sum_i \frac{M_N^{\star}(\vec{R}_i)}
{E_i-V(\vec{R}_i)}
\delta(\vec{r}-\vec{R}_i) |A \rangle, \label{mf13}\\
\rho_B(\vec{r})&=& \langle A|\sum_i \delta(\vec{r}-\vec{R}_i) |A \rangle 
,\label{mf13'}\\
\rho_3(\vec{r})&=& \langle A|\sum_i 
\frac{\tau^N_3}{2}\delta(\vec{r}-\vec{R}_i) |A \rangle ,
\label{mf13''}
\en
where we have used Eq.(\ref{46.3}) to eliminate the factor 
$(\cosh\xi_i)^{-1}$. Note that the definition of the scalar density 
makes sense because, in mean field approximation, each nucleon is moving in 
an orbital with a given energy.  The meson sources then take the simple 
form
\bg
\langle A|\overline{q}q (t, \vec{r})|A \rangle &=& 3 
S(\vec{r})\rho_s(\vec{r}),\label{mf14}\\
\langle A|\overline{q}\gamma^\nu q (t, \vec{r})|A \rangle &=& 
3 \delta(\nu,0) \rho_B(\vec{r}),\label{mf14'}\\
\langle A|\overline{q}\gamma^\nu \frac{\tau^\alpha}{2} 
q (t, \vec{r})|A \rangle 
&=& \delta(\nu,0)\delta(\alpha,3)\rho_3(\vec{r}),
\label{mf14''}
\en
where we have deduced the source of the $\rho$ from that for the $\omega$. 
We have used the notation
\bg
S(\vec{r})=S(\vec{0},\vec{r})&=&\int d\vec{u}\;s_{\vec{r}}(\vec{u}),
\label{mf15}\\
 &=& \frac{\Omega_0/2 + m_q^{\star}R_B(\Omega_0 -1)}{\Omega_0(\Omega_0-1) + 
m_q^{\star}R_B/2}, \label{mf15'}
\en
where the subscript $\vec{r}$ reminds us 
that the function $s_{\vec{r}}(\vec{u})$ must be 
evaluated in the scalar field existing at $\vec{r}$.

Since their sources are time independent and since they do not propagate, 
the mean meson fields are also time independent. So by combining 
Eqs.(\ref{mf1}) to (\ref{mf2''}) and Eqs.(\ref{mf14}), (\ref{mf14'}) and 
(\ref{mf14''}),
we get the desired equations for $\sigma(\vec{r}), \omega(\vec{r})$ and 
$b(\vec{r})$: 
\bg
(-\nabla^2_r+m^2_\sigma)\sigma(\vec{r})&=& g_\sigma C(\vec{r}) 
\rho_s(\vec{r}),\label{mf16}\\
(-\nabla^2_r+m^2_\omega)\omega(\vec{r})&=&  g_\omega  \rho_B(\vec{r}),
\label{mf16'}\\
(-\nabla^2_r+m^2_\rho)b(\vec{r})&=& g_\rho  \rho_3(\vec{r}).
\label{mf16''}
\en
where the nucleon coupling constants and $C$ are defined  by
\bge
g_\sigma =3 g_\sigma^q S(\sigma=0),\ \ g_\omega = 3 g_\omega^q
,\ \ g_\rho = g_\rho^q,\ \ C(\vec{r})=S(\vec{r})/S(\sigma=0). 
\label{mf16a}
\ene
For completeness we recall that the mean fields carry the following energy
\bge
E^{meson}=\frac{1}{2}\int d\vec{r} \, [(\vec{\nabla}\sigma)^2+
m^2_\sigma\sigma^2-(\vec{\nabla}\omega)^2-m^2_\omega\omega^2-
(\vec{\nabla}b)^2-m^2_\rho b^2].
\label{mf17}\ene

\section{The problem of self-consistency }
\label{sec:self}

Our quark model for finite nuclei is now complete. As the main equations 
are scattered through the text,  it is  useful to summarize the
procedure which we have developed as follows:
\begin{enumerate}
\item Choose the bare quark mass, $m_q$, and adjust the bag parameters, 
$B$ and $z_0$, to fit the free nucleon mass 
and its bag radius (see Eq.(\ref{42'}) and Table~\ref{b,z}).
\item Assume that the coupling constants and the masses of the mesons are 
known (see Table~\ref{c.c.} in Sec.~\ref{ssec:matter}).
\item Evaluate the nucleon properties, $I(\sigma)$ (Eq.(\ref{Idef2})) and 
$S(\sigma)$ (Eq.(\ref{mf15'})), for a range of values of $\sigma$ 
(see also Eqs.(\ref{paramC}) and (\ref{paramI}) in Sec.~\ref{ssec:matter}).
\item Guess an initial form for the densities, $\rho_s(\vec{r}), 
\rho_B(\vec{r})$ and 
$\rho_3(\vec{r})$, in Eqs.(\ref{mf13}), (\ref{mf13'}) and (\ref{mf13''}).
\item For $\rho_s(\vec{r})$ fixed, solve Eq.(\ref{mf16}) 
for the $\sigma$ field. 
\item For $\rho_B(\vec{r})$ and $\rho_3(\vec{r})$ fixed, 
solve Eqs.(\ref{mf16'}) and 
(\ref{mf16''}) for the $\omega$ and $\rho$ fields. 
\item Evaluate the effective mass $M_N^{\star}(\vec{r})$  and the potential 
$V(\vec{r})$ according to Eqs.(\ref{42'}) and (\ref{VR}). 
The bag radius at each 
point in the nucleus is fixed by Eq.(\ref{42''}). For 
practical purposes it is useful to note that  $M_N^{\star}(\vec{r})$ 
and $C(\vec{r})$ 
depend only on  the value of the $\sigma$ field at $\vec{r}$ and that  
a simple parametrization, linear in $g_{\sigma} \sigma$, works extremely 
well at moderate densities
(see Eqs.(\ref{paramC}) and (\ref{mstaR}) in Sec.~\ref{ssec:matter}). 
So  one does not need to solve the bag equations at each point when 
solving the self-consistent nuclear problem. 
\item Solve the eigenvalue problem defined by the nuclear Hamiltonian 
Eq.(\ref{nuc9}) and generate the shell model 
from which the densities $\rho_s(\vec{r}), \rho_B(\vec{r})$ and 
$\rho_3(\vec{r})$ 
can be computed according to Eqs.(\ref{mf13}), (\ref{mf13'}) and 
(\ref{mf13''}).
\item Go to 5 and iterate until self-consistency is achieved.
\end{enumerate}
This procedure has to be repeated for each nucleus, which certainly implies 
considerable numerical work. We plan to study the implications of 
this non-relativistic formulation in a future work.  

\section{Relativistic formulation}
\label{sec:relf}

Here we attempt to formulate the model as a relativistic field theory for 
the nucleon in order to have a direct comparison with the widely used QHD. 
We make no attempt to justify the formulation of a local relativistic field 
theory at a fundamental level because this is not possible for a composite 
nucleon. Our point is that, in the mean field approximation, QHD has 
had considerable phenomenological success. 
We therefore try to express our results in this 
framework. The idea is to write a relativistic Lagrangian and to check 
that, in some approximation, it is equivalent to  our non-relativistic 
formulation. 

As shown earlier, our basic result is that essentially the nucleon in the 
meson fields behaves as a point like particle of effective mass 
$M_N^{\star}(\sigma(\vec{r}))$ moving in a potential 
$g_\omega \omega(\vec{r})$. To simplify  we do not consider the $\rho$ 
coupling and therefore we shall 
only apply the model to $N=Z$  nuclei.

As already pointed out the 
 spin-orbit force  
of the $\omega$ is almost the same as the one of a point like Dirac particle. 
So a possible Lagrangian density for this system is 
\bge
{\cal L}=i\overline{\psi}\gamma \cdot \partial\psi 
-M_N^{\star}({\hat{\sigma}})\overline{\psi}\psi
-g_\omega\hat{\omega}^\mu\overline{\psi}\gamma_\mu\psi+{\cal L}_{mesons},
\label{relat1} 
\ene
where $\psi$, $\hat{\sigma}$ and $\hat{\omega}^\mu$ are respectively  the 
nucleon, $\sigma$ and $\omega$ field operators. 
The free meson Lagrangian density is 
\bge
{\cal L}_{mesons}=\frac{1}{2}(\partial_\mu\hat{\sigma}\partial^\mu 
\hat{\sigma}-
m_{\sigma}^2\hat{\sigma}^2)-\frac{1}{2}\partial_\mu\hat{\omega}_\nu
(\partial^\mu\hat{\omega}^\nu-\partial^\nu\hat{\omega}^\mu)+
\frac{1}{2}m_\omega^2\hat{\omega}^\mu\hat{\omega}_\mu .
\label{relat2}
\ene
The comparison with QHD is straightforward. If we define the field 
dependent coupling constant $g_\sigma(\hat{\sigma})$ by
\bge
M_N^{\star}(\hat{\sigma})=M_N-g_\sigma(\hat{\sigma})\hat{\sigma}, 
\label{relat6}
\ene
it is easy to check that $g_\sigma(\sigma=0)$ is equal to the coupling 
constant $g_\sigma$ defined in Eq.(\ref{mf16a}). Then we  write the 
Lagrangian density as 
\bge
{\cal L}=i\overline{\psi}\gamma \cdot \partial\psi- M_N\overline{\psi}\psi
+g_\sigma(\hat{\sigma})\hat{\sigma}\overline{\psi}\psi 
-g_\omega\hat{\omega}^\mu\overline{\psi}\gamma_\mu\psi+{\cal L}_{mesons},
\label{relat7} 
\ene
and clearly the only difference from QHD lies in the fact that the internal 
structure of the nucleon has forced a  {\it known} dependence of the scalar 
meson-nucleon coupling constant on the scalar field itself.  Note that this 
dependence is not the same as the one adopted in the density dependent 
hadron field theory~\cite{fuchs} where the meson-nucleon vertices are 
assumed to depend directly on the baryonic densities.  

In the mean field approximation, the meson field operators are replaced by 
their time independent expectation values in the ground state of the nucleus: 
\bge
\hat{\sigma}(t,\vec{r})\to\sigma(\vec{r}),\ \ \ 
\hat{\omega}^\mu(t,\vec{r})\to\delta(\mu,0)\omega(\vec{r}),
\label{relat3}
\ene
and variation of the Lagrangian  yields the Dirac equation
\bge
( i\gamma \cdot \partial -M_N^{\star}(\sigma)- 
g_\omega \gamma_0 \omega ) \psi = 0 ,
\label{relat4}
\ene
as well as the equations for the meson mean fields
\bg
(-\nabla^2_r+m^2_\sigma)\sigma(\vec{r})&=& - \left( 
\frac{\partial}{\partial\sigma}M_N^{\star}(\sigma) \right) \langle 
A|\overline{\psi}\psi(\vec{r})|A \rangle, \label{relat5.1}\\
(-\nabla^2_r+m^2_\omega)\omega(\vec{r})&=&  
g_\omega \langle A|\psi^\dagger\psi(\vec{r})|A \rangle .
\label{relat5.2}
\en

Using the Foldy-Wouthuysen transformation (see Ref.\cite{larry} for 
the details), one can show that the Dirac equation (\ref{relat4}) gives 
back our non-relativistic, quantum Hamiltonian (without the $\rho$ coupling)
under the following conditions:
\begin{itemize}
\item only terms of second order in the velocity are kept,
\item second derivatives of the meson fields are ignored,
\item the fields are small with respect to the nucleon mass,
\item the difference between  $\mu_s=.88$ and $\mu_s(point)=1$ can be 
neglected. 
\end{itemize}

 The fact that $\mu_v=4.7$ is very different from $\mu_v(point)=1$ prevents 
one from using this simple scheme  for the $\rho$ coupling. 
The spin-orbit potential would be completely different from the one 
obtained in Eq.(\ref{46.06}). Since the $\rho$ is not required in the 
treatment 
of symmetric nuclei we postpone to future work its introduction in 
the relativistic framework.

In the same approximations one finds that $\langle 
A|\overline{\psi}\psi(\vec{r})|A \rangle$ 
and $\langle A|\psi^\dagger\psi(\vec{r})|A \rangle$ are respectively 
equal to the previously defined scalar source, $\rho_s$ (Eq.(\ref{mf13})), 
and vector source, $\rho_B$ (Eq.(\ref{mf13'})). 
(In nuclear matter the approximation is in fact exact). 
Therefore the equations (\ref{relat5.1}) and (\ref{relat5.2}) for the 
meson fields reproduce our previous results given in  
Eqs.(\ref{mf16}) and (\ref{mf16'}) provided the relation
\bg
C(\sigma)g_\sigma(\sigma=0)&=&
-\frac{\partial}{\partial\sigma}M_N^{\star}(\sigma) , \\
&=&\frac{\partial}{\partial\sigma}(g_\sigma(\sigma)\sigma) , 
\label{relat10}
\en 
is satisfied, which can be checked explicitly from the definition 
of $M_N^{\star}(\sigma)$ and $C(\sigma)$.

Within the limits specified above, the Lagrangian density given in 
Eq.(\ref{relat1}) looks acceptable and we shall proceed to study its
numerical consequences.

\section{Applications}
\label{sec:appl}

Before turning to finite nuclei, we first explain briefly how the
general formalism applies in the case of nuclear matter.

\subsection{Infinite Nuclear Matter}
\label{ssec:matter}

In symmetric, infinite nuclear matter the sources of the fields 
are constant and can be related
to the nucleon Fermi momentum $k_F$ according to~\cite{serot}

\begin{figure}[htb]
\centering{\
\epsfig{file=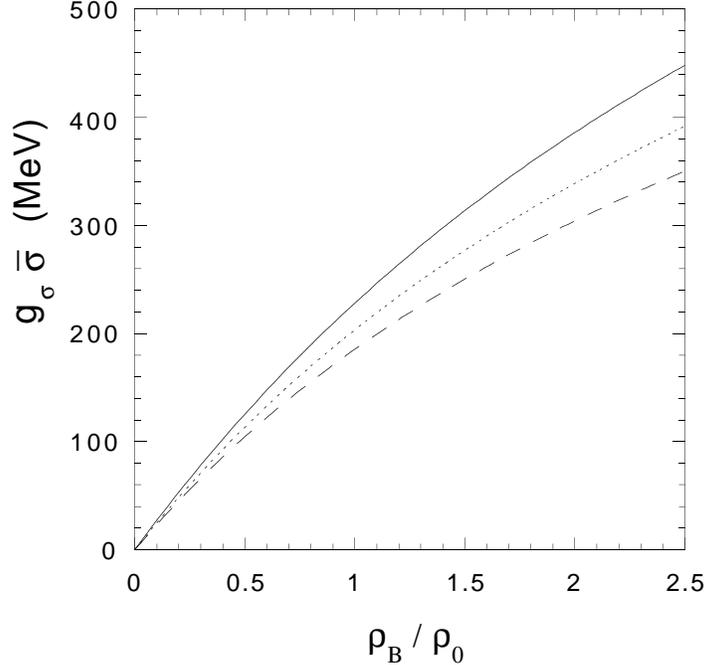,height=9cm}
\caption{Mean-field values of the $\sigma$ meson for various bag radii
as a function of $\rho_B$.
The solid, dotted and dashed
curves show $g_{\sigma} {\overline \sigma}$ for $R_B^0$ = 0.6, 0.8 and
1.0 fm, respectively. The quark mass is chosen to be 5 MeV.}
\label{fig:sigma}}
\end{figure}
\begin{figure}[hbt]
\centering{\
\epsfig{file=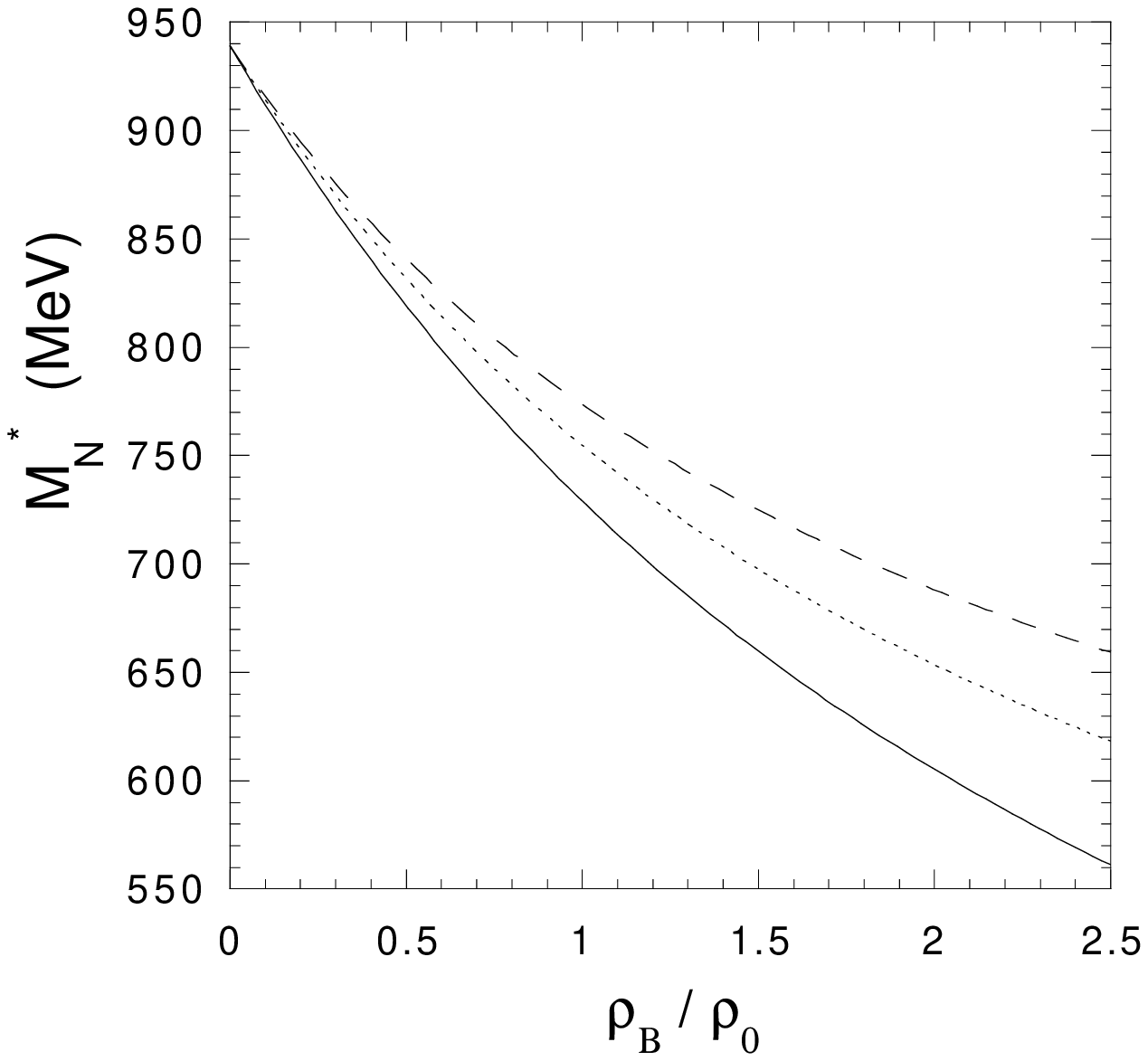,height=9cm}
\caption{Effective nucleon mass ($m_q$=5MeV). The curves are
labelled as in Fig.1. }
\label{fig:mass}}
\end{figure}
\bg
\langle A|\psi^\dagger\psi(\vec{r})|A \rangle &=& 
\frac{4}{(2\pi)^3}\int^{k_F}d\vec{k}=\frac{2 k_F^3}{3\pi^2} , 
\label{self1}\\
\langle A|\overline{\psi}\psi(\vec{r})|A \rangle &=& 
\frac{4}{(2\pi)^3}\int^{k_F}  
d\vec{k}\frac{M_N^{\star}}{\sqrt{M_N^{\star 2}+\vec{k}^2}},
\label{self1'}
\en
where $M_N^{\star}$ denotes the constant value of the effective nucleon 
mass defined 
by Eq.(\ref{42'}), or equivalently, Eq.(\ref{relat6}). 
Obviously these equations can also be deduced 
from the definitions (Eqs.(\ref{mf13}) and (\ref{mf13'})) of the scalar and 
vector densities provided that, for  $\rho_s$,  one  uses 
Eqs.(\ref{46.3}) and (\ref{46.4}) to write
\bge
\frac{M_N^{\star}}{E-V}=\frac{1}{\cosh\xi}=\frac{M_N^{\star}}
{\sqrt{M_N^{\star 2}+\vec{k}^2}}.
\label{self2}
\ene
\begin{figure}[htb]
\centering{\
\epsfig{file=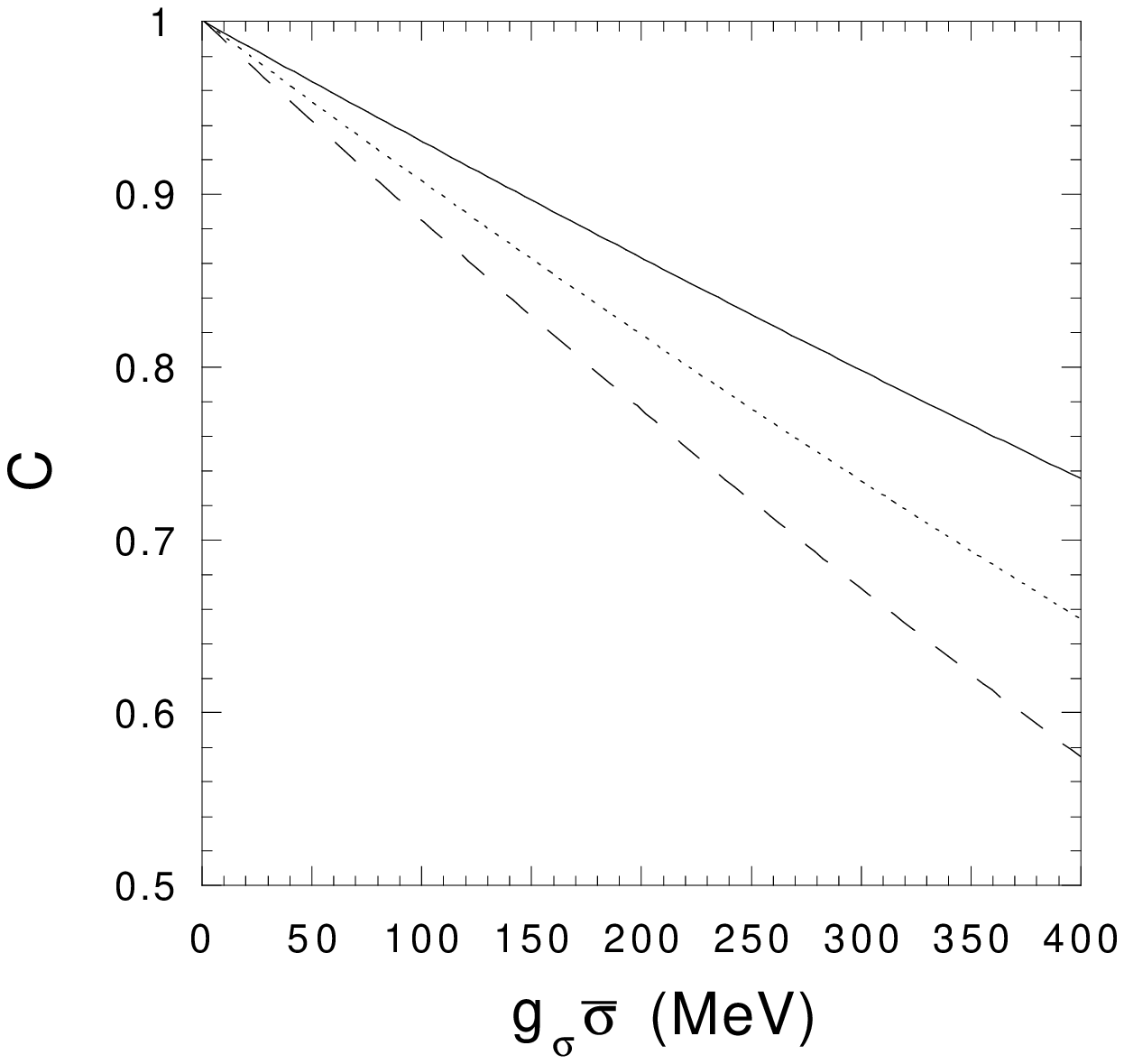,height=9cm}
\caption{Scalar density ratio, $C({\overline \sigma})$, as a function of
$g_{\sigma} {\overline \sigma}$ ($m_q$ = 5 MeV). The curves are
labelled as in Fig.1. }
\label{fig:C}}
\end{figure}
\begin{figure}[bt]
\centering{\
\epsfig{file=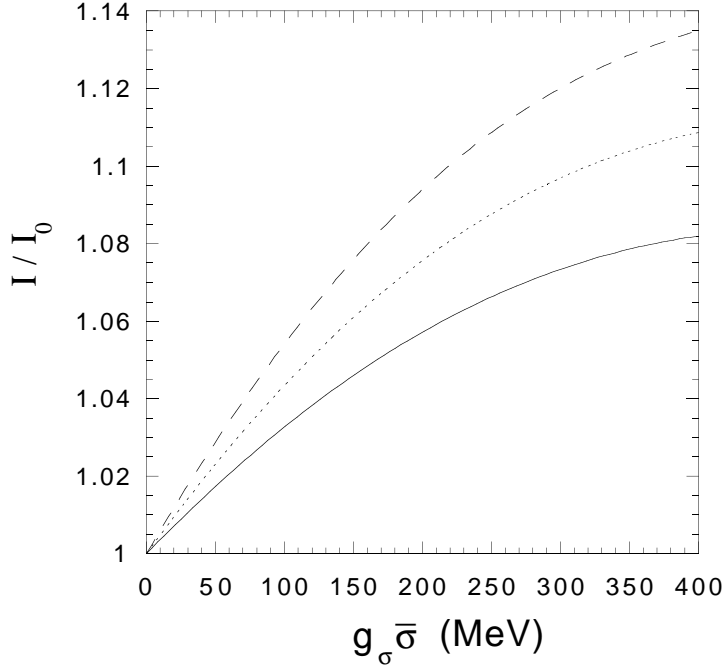,height=9cm}
\caption{The ratio of $I({\overline \sigma})$ to $I_0$ ($m_q$ = 5 MeV).
The curves are labelled as in Fig.1. }
\label{fig:I}}
\end{figure}

Let ($\overline{\sigma}, \overline{\omega}$) be the 
constant mean-values of the meson fields. From 
Eqs.(\ref{mf16}) and (\ref{mf16'})
we find
\bg
\overline{\omega}&=&\frac{g_\omega \rho_B}{m_\omega^2},\label{self3}\\
\overline{\sigma}&=&\frac{g_\sigma }{m_\sigma^2}C(\overline{\sigma})
\frac{4}{(2\pi)^3}\int^{k_F} 
d\vec{k}\frac{M_N^{\star}}{\sqrt{M_N^{\star 2}+\vec{k}^2}},\label{self3'}
\en
where $C(\overline{\sigma})$ is now the constant value of $C$  in the 
scalar field.\footnote{Note the change
in notation from the earlier papers of Saito and Thomas
where $C$ was used for 
what we now call $S$.}
As emphasised by 
Saito and Thomas~\cite{st1}, the
self-consistency equation for $\overline{\sigma}$, Eq.(\ref{self3'}), 
is the same as that in QHD {\it except} that in the 
latter model one has $C(\overline{\sigma})=1$ 
(i.e. the quark mass is infinitely heavy). 

Once the self-consistency equation for $\overline{\sigma}$ has been solved, 
one can evaluate the energy per nucleon. 
From the Dirac equation (\ref{relat4}), or simply using Eq.(\ref{46.3}), 
the energy of a nucleon with momentum $\vec{k}$ is
\bge
E(\vec{k})=V+M_N^{\star}\cosh\xi=g_\omega\overline{\omega}+\sqrt{M_N^{\star 
2}+\vec{k}^2}, 
\label{self4}
\ene 
which contributes to the  energy per nucleon  by the amount
\bg
E^{nucl.}/A&=& \frac{4}{\rho_B (2\pi)^3}\int^{k_F} d\vec{k}E(\vec{k}), 
\label{self5}\\
&=&\frac{1}{\rho_B}\left[g_\omega\overline{\omega}\rho_B+
\frac{4}{(2\pi)^3}\int^{k_F}d\vec{k}\sqrt{M_N^{\star 2}+\vec{k}^2}\right].
\label{self5'}
\en
The contribution of the energy stored in the meson fields is, from 
Eq.(\ref{mf17}), 
\bge
E^{meson}/A=\frac{1}{2\rho_B}(m_\sigma^2\overline{\sigma}^2-m_\omega^2
\overline{\omega}^2),
\label{self6}
\ene
and using the expression for $\overline{\omega}$ we 
finally obtain the following expression for the  energy per nucleon
\bge
E^{total}/A=\frac{1}{\rho_B}\left[\frac{4}{(2\pi)^3}\int^{k_F}
d\vec{k}\sqrt{M_N^{\star 2}+
\vec{k}^2}+\frac{m_\sigma^2\overline{\sigma}^2}{2}+
\frac{g_\omega^2\rho_B^2}{2m_\omega^2}\right].
\label{self7}
\ene

We determine the coupling constants, 
$g_{\sigma}$ and $g_{\omega}$, so as to fit the binding 
energy ($-15.7$ MeV) per nucleon and 
the saturation density ($\rho_0$ = 0.15 fm$^{-3}$) for 
symmetric nuclear matter at equilibrium. 
The coupling constants and some calculated properties
of nuclear matter (with $m_q$ = 5 MeV, $m_{\sigma}=550$ MeV and 
$m_{\omega}=783$ MeV) at the 
saturation density are listed in Table~\ref{c.c.}. 
The most notable fact is that the calculated incompressibility, $K$, 
is well within the experimental range: $K \approx 200 - 
300$ MeV~\cite{compress}. Also our effective nucleon mass 
is much larger than in the case of QHD. 
In the last two columns of Table~\ref{c.c.} we show the 
relative modifications (with respect to their values at zero density) 
of the bag radius and the lowest eigenvalue, $x$, at 
saturation density.  The changes are not large.  In order to show the
relative insensitivity to the quark mass (as long as it is small) 
we note that 
$M_N^{\star}$ = (756, 753) MeV at saturation density and $K$ = 
(278, 281) MeV for $R_B^0$ = 0.8 fm and $m_q$ = (0, 10) MeV, respectively. 
\begin{table}[hbtp]
\begin{center}
\caption{Coupling constants and calculated nucleon properties in
symmetric nuclear matter at normal nuclear matter density.  
The effective nucleon mass, 
$M_N^{\star}$, and the 
nuclear incompressibility,  
$K$, are quoted in MeV. The bottom row is for QHD.}
\label{c.c.}
\begin{tabular}[t]{cccccccc}
\hline
 $R_B^0$(fm)&$g_{\sigma}^2/4\pi$&$g_{\omega}^2/4\pi$&$M_N^{\star}$&$K$&
$\frac{\delta R_B}{R_B^0}$&$\frac{\delta x}{x_0}$ \\
\hline
  0.6 & 5.86 & 6.34 & 729 & 295 & -0.02 & -0.13 \\
  0.8 & 5.40 & 5.31 & 754 & 280 & -0.02 & -0.16 \\
  1.0 & 5.07 & 4.56 & 773 & 267 & -0.02 & -0.21 \\
  QHD & 7.29 & 10.8 & 522 & 540 & --    & --    \\
\hline
\end{tabular}
\end{center}
\end{table}

In Figs.\ref{fig:sigma} and \ref{fig:mass}, we show the mean-field values 
of the $\sigma$ meson and the effective nucleon mass in medium, respectively. 
In both cases their dependence on the bag radius is rather weak. 
The scalar density ratio, $C({\overline \sigma})$, and the ratio of 
the integral 
$I({\overline \sigma})$ to $I_0$ are plotted as a function of 
$g_{\sigma} {\overline \sigma}$ in Figs.\ref{fig:C} and \ref{fig:I}, 
respectively.  
As ${\overline \sigma}$ increases, 
the scalar density ratio decreases linearly, 
while the ratio, $I/I_0$, gradually increases.  Here we note that $S(0)$ = 
(0.4819, 0.4827, 0.4834) and $I_0$ = (0.2421, 0.3226, 0.4028) fm for 
$R_B^0$ = (0.6, 0.8, 1.0) fm, respectively.  

It would be very useful to give a simple parametrization for $C$ 
and $I/I_0$ 
because they are completely controlled by only the strength of the local 
$\sigma$ field.  We can easily see that $C$ is well approximated  
by the linear form: 
\bge
C({\overline \sigma}) = 1 - a \times (g_{\sigma}{\overline \sigma}) , 
\label{paramC}
\ene
with $g_{\sigma} {\overline \sigma}$ in MeV and 
$a$ = (6.6, 8.8, 11) $\times 10^{-4}$, for $R_B^0$ = (0.6, 0.8, 1.0) fm, 
respectively.  
For $I/I_0$ we find a quadratic form: 
\bge
\frac{I({\overline \sigma})}{I_0} = 1 + b_1 \times 
(g_{\sigma}{\overline \sigma}) - b_2 \times 
(g_{\sigma} {\overline \sigma})^2 , 
\label{paramI}
\ene
with $b_1$ = (3.7, 4.9, 6.1) $\times 10^{-4}$ and $b_2$ = (3.9, 5.2, 6.5) 
$\times 10^{-7}$.  
More comments and discussion of the results for nuclear matter
can be found in the previous 
publications~\cite{guichon,st1}.

As a practical matter, we note that Eq.(\ref{relat10}) is easily solved for
$g_{\sigma}(\sigma)$ in the case where $C(\sigma)$ is linear in $g_{\sigma} 
\bar{\sigma}$  -- as we found in Eq.(\ref{paramC}). In fact, it is easy to 
show 
that
\bge
M^{\star}_N = M_N - \left[ 1 - \frac{a}{2} (g_\sigma {\overline \sigma}) 
\right] (g_\sigma {\overline \sigma}) , 
\label{mstaR}
\ene
(recall $g_{\sigma} \equiv g_{\sigma}(\sigma=0)$, Eq.(\ref{mf16a})) so that
the effective $\sigma$N coupling constant decreases at half the rate 
of $C(\sigma)$. (Equation (\ref{mstaR}) is quite accurate up to  twice 
nuclear matter density.) Having explicitly solved the nuclear matter problem 
by 
self-consistently solving for the quark wave functions in the bag in the 
mean scalar field {\it one can solve for the properties of finite nuclei 
without explicit reference} to the internal structure of the nucleon. 
All one needs is Eqs.(\ref{paramC}) and
(\ref{mstaR}) for $C(\sigma)$ and $M^{\star}_N$ as a function 
of $g_{\sigma} \bar{\sigma}$.

\clearpage

\subsection{Initial results for finite nuclei}
\label{ssec:finite}

In order to keep the present work to a reasonable length we  
propose to present detailed numerical studies of the properties of
finite nuclei calculated within the present framework in a later
paper~\cite{later}. However, to illustrate that the approach
has some promise, we have performed some preliminary calculations for the
doubly closed shell nucleus $^{16}O$. This requires the self-consistent
solution of our equations for 6 Dirac orbitals -- the $1s_{1/2}$, 
$1p_{1/2}$ and $1p_{3/2}$ 
states of the protons and neutrons. The central Coulomb interaction has been 
taken into account for the protons. 

\begin{figure}[htb]
\centering{\
\epsfig{angle=270,figure=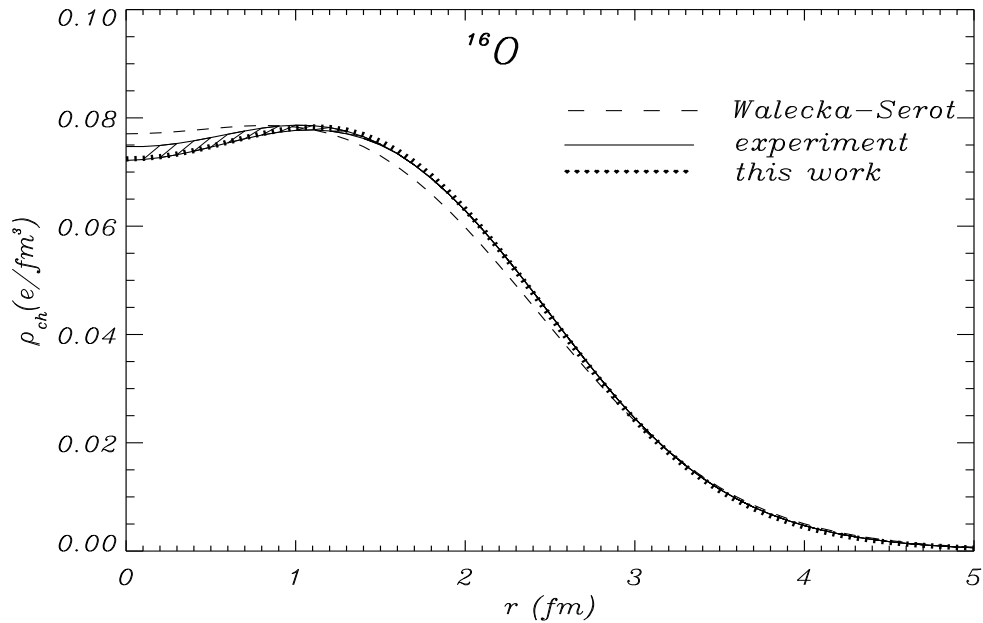,height=9cm,width=14cm,angle=0}
}
\parbox{130mm}{\caption{The charge density of $^{16}O$ in the 
present model and QHD, compared with the experimental distribution.}
\label{fig:oxygen}}
\end{figure}

\begin{figure}[htb]
\centering{\
\epsfig{angle=270,figure=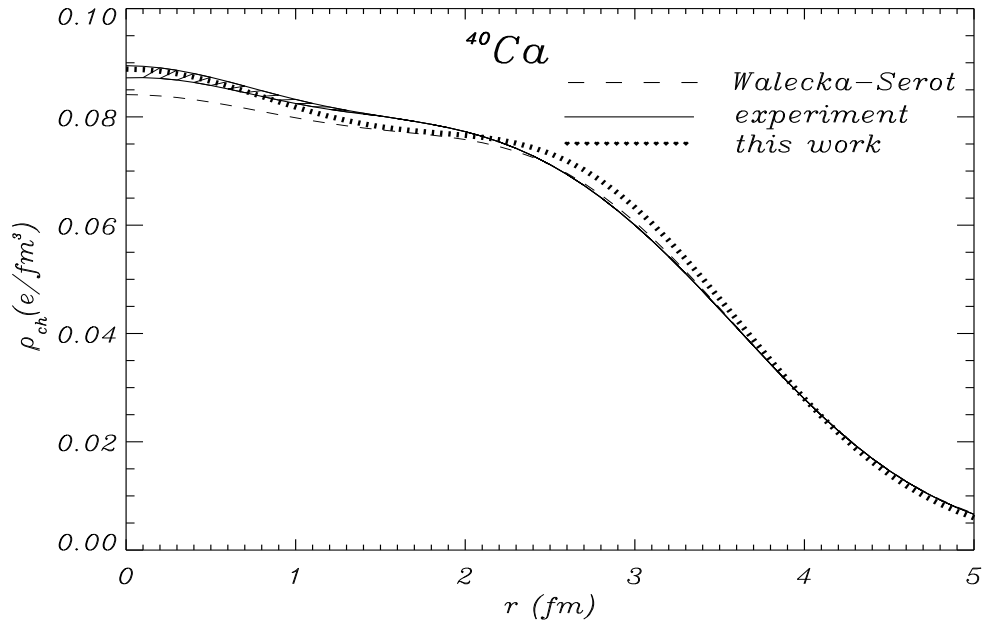,height=9cm,width=14cm,angle=0}
}
\parbox{130mm}{\caption{The charge density of $^{40}Ca$ in the 
present model and QHD, compared with the experimental distribution.}
\label{fig:calcium}}
\end{figure}

The numerical calculation was 
carried out using the techniques described by Walecka and Serot~\cite{serot}.
The resulting charge density for $^{16}O$ is shown in Fig.\ref{fig:oxygen}
(dotted curve) in comparison with the experimental 
data~\cite{oxydata} (hatched area) and QHD~\cite{serot}.
\begin{figure}[htb]
\centering{\
\epsfig{angle=270,figure=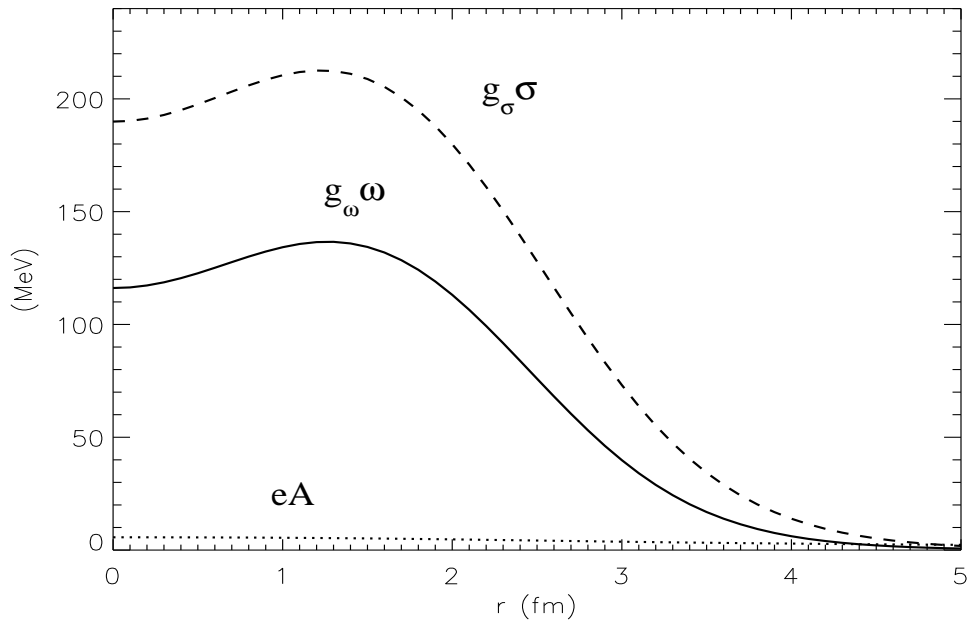,height=9cm,width=14cm,angle=0}
}
\parbox{130mm}{\caption{Scalar, vector and coulomb potentials for 
$^{16}O$ in the present model.}
\label{fig:meson}}
\end{figure}
In this
calculation we used the parameters given in Table~\ref{c.c.} for $R_B^0 =
0.8$ fm. However, as the
central density of $^{16}O$ was a little high we increased the
model-dependent slope of the scalar density $C(\sigma)$ and the 
coupling constant $g_{\sigma}(\sigma)$ (i.e. the
parameter $a$ in Eqs.(\ref{paramC}) and (\ref{mstaR}) ) by $10\%$ 
to obtain the result
shown. The corresponding effect on the saturation energy and density of
nuclear matter is very small: $\rho_0 \rightarrow 0.1496$ fm$^{-3}$ 
and the energy per nucleon becomes $-15.65$ MeV. 
In Fig.\ref{fig:meson}, we also show the scalar and vector fields 
corresponding to the charge density of $^{16}O$ that was shown in 
Fig.\ref{fig:oxygen}.  Using the same parameter set one also finds a very 
reasonable fit to the charge density of $^{40}Ca$.  It is presented in 
Fig.\ref{fig:calcium}.  

To conclude this initial investigation of finite nuclei we record, in
Table~\ref{s.o.}, the calculated single particle binding energies of the
protons and neutrons in 
$^{40}Ca$, in comparison with the results of QHD and the experimental 
data~\cite{boer}.  
\newcommand{\lw}[1]{\smash{\lower2.ex\hbox{#1}}}
\begin{table}[hbtp]
\begin{center}
\caption{Predicted proton and neutron spectra of $^{40}Ca$ compared with 
QHD and the experimental data.  QMC(5, 300) means the 
present model with $R_0=$ 
0.8 fm and $m_q=$ 5 and 300 MeV, respectively (see text). All energies are in
MeV. }
\label{s.o.}
\begin{tabular}[t]{ccccccccc}
\hline
\lw{Shell} & \multicolumn{4}{c}{neutron} & 
\multicolumn{4}{c}{proton} \\
\cline{2-9} 
 & QHD  & QMC(5) & QMC(300) & Expt. & QHD & QMC(5) & QMC(300) & Expt. \\
\hline
$1s_{1/2}$ & 54.9 & 43.5 & 47.1 & 51.9 &  46.7 & 35.5 & 38.9 & 50$\pm$10 \\
$1p_{3/2}$ & 38.6 & 32.5 & 34.5 & 36.6 &  30.8 & 24.7 & 26.7 & 34$\pm$6 \\
$1p_{1/2}$ & 33.1 & 31.3 & 32.4	& 34.5 &  25.3 & 23.5 & 24.5 & 34$\pm$6 \\
$1d_{5/2}$ & 22.6 & 20.0 & 21.0 & 21.6 &  15.2 & 12.6 & 13.6 & 15.5 \\
$2s_{1/2}$ & 14.6 & 15.0 & 15.3 & 18.9 &   7.4 &  7.7 &  7.9 & 10.9 \\
$1d_{3/2}$ & 14.1 & 17.9 & 17.3 & 18.4 &   6.8 & 10.5 &  9.8 & 8.3  \\
\hline
\end{tabular}
\end{center}
\end{table}
Because of the smaller scalar and vector field strengths in the present model,
compared with QHD, the spin-orbit splitting tends also to be smaller -- 
perhaps only $\frac{2}{3}$ of the experimentally observed splittings.
In this context it is interesting to show the sensitivity of 
the present model to just one feature of the underlying structure of the 
nucleon, namely the mass of the confined quark. As an example, we consider 
the case $m_q = 300$ MeV, which is a typical constituent quark mass. For
$m_q=$ 300 MeV and $R_0=$ 0.8 fm, the coupling constants required to fit the 
saturation properties of nuclear matter are 
$g_{\sigma}^2/4\pi 
=$ 6.84 and $g_{\omega}^2/4\pi=$ 8.51, and the effective nucleon mass 
(at saturation) and the 
incompressibility become 674 MeV and 334 MeV, respectively. 
Using these parameters one finds that 
the charge densities of $^{16}O$ and $^{40}Ca$ are again reproduced very
well (without any need to vary the slope parameter 
``$a$", i.e., $a = 3.9 \times 10^{-4}$).  From the table, 
one can see that a heavier quark mass gives a spectrum closer to that of 
QHD -- as discussed by
Saito and Thomas in Ref.\cite{st1} -- and in better agreement 
with the observed
spin-orbit splittings.  In conclusion we notice 
one more point: the $2s_{1/2}$ and $1d_{3/2}$ levels are inverted compared 
with the experimental data (see also Ref.\cite{boer}).  
This may be connected with the effect of rearrangement~\cite{john,campi}, 
which is not considered here.  Also, the effect of antisymmetrization 
remains to be investigated.  

\clearpage

\section{Summary}
\label{sec:summ}

Starting with a hybrid model in which quarks confined in nucleon bags
interact through the exchange of scalar and vector mesons, we have shown
that the Born-Oppenheimer approximation leads naturally to a
generalisation of QHD with a density dependent scalar coupling. The
physical origin of this density dependence, which provides a new
saturation mechanism for nuclear matter, is the relatively rapid
increase of the lower Dirac component of the wavefunction of the
confined, light quark. We confirm the original discovery of
Guichon~\cite{guichon} that once the scalar and vector coupling constants
are chosen to fit the observed saturation properties of nuclear matter
the extra, internal degrees of freedom lead to an incompressibility that
is consistent with experiment.

In the case of finite nuclei we have derived a set of coupled
differential equations which must be solved self-consistently but which
are not much more difficult to solve than the relativistic Hartree
equations of QHD. Initial results for $^{16}O$ 
(see Fig.\ref{fig:oxygen}) are quite promising but a full numerical
study will be presented in a later work~\cite{later}.

The successful generalisation of the quark-meson coupling model to
finite nuclei opens a tremendous number of opportunities for further
work. For example, earlier results for the Okamoto-Nolen-Schiffer
anomaly~\cite{st3}, the nuclear EMC effect~\cite{st2}, 
the charge-symmetry violating correction to
super-allowed Fermi beta-decay~\cite{wilk}
and so on, can now be treated in a truly quantitative way.

It will be very interesting to explore the connection between the
density dependence of the variation of the effective $\sigma$-nucleon
coupling constant, which arises so naturally here, with the variation 
found empirically in earlier work. We note, in particular, that while
our numerical results depend on the particular model chosen here (namely, 
the MIT bag model), the qualitative features which we find (such as the
density dependent decrease of the scalar coupling) will apply in any
model in which the nucleon contains light quarks and the attractive $N$-$N$ 
force is a Lorentz scalar. Of course, it will be important to
investigate the degree of variation in the numerical results for other
models of nucleon structure.

We could list many other directions for future theoretical work: for
the replacement of the MIT bag by a model respecting PCAC (e.g. the
cloudy bag model~\cite{cbm}), the replacement of $\sigma$-exchange by
two-pion exchange, the replacement of $\omega$ exchange by nucleon
overlap at short distance, the inclusion of the density dependence of
the meson masses themselves~\cite{hadron,jpw} and so on. On the practical
side, we stress that the present model can be applied to all the
problems for which QHD has proven so attractive, with very little extra
effort. It will also be interesting to explore its phenomenological
consequences in this way.

\section*{Acknowledgements}
We would like to thank both Dr. B. Jennings and the unknown referee who
pointed out an error in the treatment of the spin-orbit term in an
earlier version of this paper.
This work was supported by the Australian Research Council and the
French Commissariat \`a l'Energie Atomique.

\newpage
\begin{flushleft}
\begin{Large}
{\bf Appendix}
\end{Large}
\end{flushleft}

Here we want to demonstrate that the center of mass correction to the
bag energy is essentially independent of the external scalar field.
(Since the vector fields do not alter the quark structure of the nucleon
we need not consider them.) To avoid the difficulties associated with
the confinement by a sharp boundary, we consider a model where
the quark mass grows quadratically with the distance from the center of
the bag. This is justified because we do not look for the c.m. correction
itself but only for its dependence on the external field. Moreover, 
after the strength of the confining mass has been  adjusted to 
reproduce the lowest 
eigenfrequency of the bag, we have found that the corresponding wave
functions are rather similar to those for the bag.

 To estimate the c.m. correction we also make the assumption that the quark
number is a good quantum number, which allows us to formulate the problem
in the first quantized form. This amounts to neglecting the effect of 
quark-antiquark excitation and is therefore not a very strong constraint.

Thus the model is defined by the following first-quantized Hamiltonian: 
\bge
H_B=\sum_{i=1, N}\gamma_0(i)[\vec{\gamma}(i)\cdot\vec{p}_i+m(\vec{r}_i)],
\ \ \ \vec{p}_i=-i\vec{\nabla}_i,
\label{appen1}
\ene
with
\bge
m(\vec{r})=m^{\star}+Kr^2,
\label{appen2}
\ene
where $m^{\star}$ is the mass of the quark in the presence of the external 
scalar field.

By assumption, $N$ is a number, so we can define intrinsic coordinates 
$({\vec \rho}, {\vec \pi})$ by
\bg
\vec{\pi}_i&=&\vec{p}_i-\frac{\vec{P}}{N},\ \ \vec{P}=\sum_i\vec{p}_i,\ \ 
\sum_i\vec{\pi}_i=0,\\
\vec{\rho}_i&=&\vec{r}_i-\vec{R},\ \ \vec{R}=\frac{1}{N}\sum_i\vec{r}_i,\ \ 
\sum_i\vec{\rho}_i=0.
\label{appen3}
\en
Then we can write the Hamiltonian in the form:
\bg
H_B&=&H_{intr.}+H_{CM},\\
H_{intr.}&=&\sum_i\gamma_0(i)[\vec{\gamma}(i)\cdot\vec{\pi}_i+
m(\vec{\rho}_i)], \\
H_{CM}&=&\frac{\vec{P}}{N}\cdot\sum_i\gamma_0(i)\vec{\gamma}(i)+
\sum_i\gamma_0(i)[m(\vec{r}_i)-m(\vec{\rho}_i)].
\label{appen4}
\en
This separation into an intrinsic and a c.m. Hamiltonian is correct because:
\begin{enumerate}
\item $H_{intr.}$ commutes with $\vec{P}$ and $\vec{R}$,
\item One has $[H_{CM}, \vec{R}]=-i\sum_i\gamma_0(i)\vec{\gamma}(i)$. Since 
for a Dirac particle $\gamma_0\vec{\gamma}$ is the velocity, 
one can identify the RHS of the previous equation with the time 
derivative of $\vec{R}$, which is consistent.
\end{enumerate}
The fact that the c.m. Hamiltonian depends on the intrinsic 
coordinates
is not a surprise because the separation is only complete 
in certain special cases.

We now look for the intrinsic energy of the bag, writing
\bge
H_{intr.}=H_B-H_{CM}.
\label{appen5}
\ene
All that we know are the eigenstates of $H_B$ but we can consider $H_{CM}$ 
as a correction of order $1/N$  with respect to the leading term in the 
bag energy. Therefore we estimate its effect in first order 
perturbation theory, that is
\bge
E_{intr.}=E_B- \langle B|H_{CM}|B \rangle =E_B-E_{CM},\label{appen6}
\ene
where $|B\rangle$ is the eigenstate of $H_B$ with energy $E_B$.  
We must therefore evaluate
\bg
E_{CM}&=&\langle B|\frac{\vec{P}}{N}\cdot\sum_j\gamma_0(j)\vec{\gamma}(j)+
\sum_i\gamma_0(i)\left[ m(\vec{r}_i)-m(\vec{r}_i-\vec{R})\right]|B\rangle 
\nonumber\\
&=&\langle B|\frac{\vec{P}}{N}\cdot\sum_j\gamma_0(j)\vec{\gamma}(j)+
2K\vec{R}\cdot\sum_i\gamma_0(i)\vec{r}_i-KR^2\sum_i\gamma_0(i)|B\rangle .
\label{appen7}
\en
 (Note that there are no ordering problems as long as $N$ is a number). 
Let $|\alpha\rangle$ be the one body solutions, that is (in units such 
that $R_B=1$)
\bg
\gamma_0[\vec{\gamma}\cdot\vec{p}+m(\vec{r})]\phi_\alpha= 
\Omega_\alpha\phi_\alpha. 
\label{appen8}
\en
If we  assume that $|B\rangle$ has all the quarks in the lowest mode then 
elementary techniques for many-body systems lead to the results
\bg
\langle B|\frac{\vec{P}}{N}\cdot\sum_j\gamma_0(j)\vec{\gamma}(j)|B\rangle 
&=&\Omega_0-\langle 0|\gamma_0(m^{\star}+Kr^2)|0\rangle ,\nonumber\\
\langle B|\vec{R}\cdot\sum_i\gamma_0(i)\vec{r}_i|B\rangle 
&=&\langle 0|\gamma_0r^2|0\rangle ,\nonumber\\
\langle B|R^2\sum_i\gamma_0(i)|B\rangle 
&=&\frac{1}{N}\langle 0|\gamma_0r^2|0\rangle 
+\left(1-\frac{1}{N}\right)\langle 0|\gamma_0|0\rangle \langle 
0|r^2|0\rangle .\nonumber\\ 
\label{appen9}
\en
so that we get
\bg
E_{CM}&=&\Omega_0-m^{\star}\langle 0|\gamma_0|0\rangle 
+K \left(1-\frac{1}{N}\right)\left(\langle 0|\gamma_0r^2|0\rangle -
\langle 0|\gamma_0|0\rangle \langle 0|r^2|0 \rangle \right)\nonumber\\
&=&\Omega_0-m^{\star}\langle 0|\gamma_0|0\rangle 
+K \left(\langle 0|\gamma_0r^2|0\rangle -\langle 
0|\gamma_0|0\rangle \langle 0|r^2|0 \rangle \right)+ 
{\cal O}(1/N). \nonumber\\ 
\label{appen10}
\en
Note that we keep only the leading term in $1/N$. This is  consistent 
with our initial approximation
according to which $E_{CM}$ is computed as a correction of order $1/N$
 with respect to the leading term in the bag energy.
\begin{figure}[tbh]
\centering{\
\epsfig{file=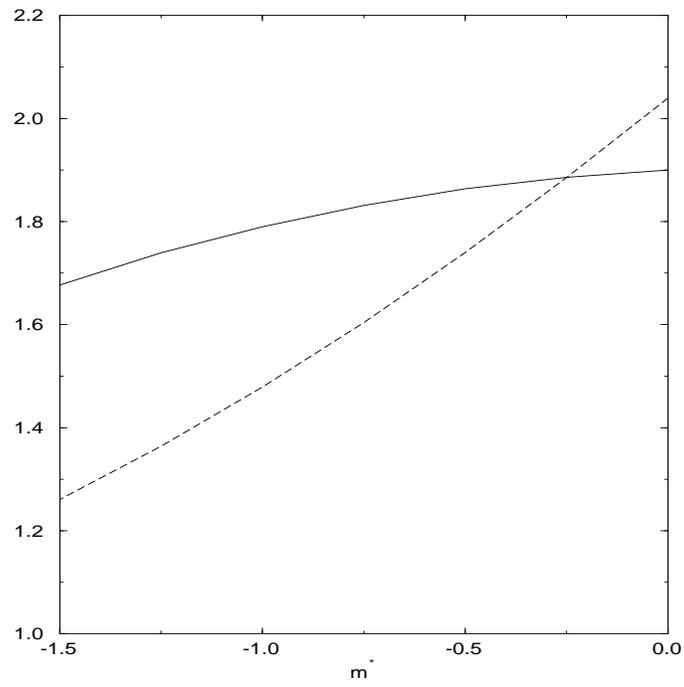,height=9cm,width=9cm}
\caption{Dependence  of $E_{CM}$ (full line) and $\Omega_0$ (dashed line) 
on $m^{\star}$.}
\label{fig:Ecm} 
}
\end{figure}

To proceed we need to evaluate the single particle matrix elements which
appear in the expression for $E_{CM}$. To determine the wave function we 
solve Eq.(\ref{appen8}) numerically and adjust the constant $K$ 
to give $\Omega_0=2.04$ -- i.e. the lowest energy level of the free bag, in
units such that $R_B=1$.  We found $K=1.74$.

Then we compute $E_{CM}$ numerically according to Eq.(\ref{appen10}), 
as a function of $m^{\star}$. 
The result is shown in Fig.\ref{fig:Ecm}, where we also plot the value of 
$\Omega_0$. One can
see that in the range $-1.5<m^{\star}<0$, which certainly contains the 
possible values of $m^{\star}$
in the case of finite real nuclei, the value of $E_{CM}$
is almost constant. For instance at $m^{\star}=-1$, $E_{CM}$ 
differs from its free value by only 6\%. 
Furthermore, its variation is clearly negligible with respect to that
of $\Omega_0$. Thus, for
practical purposes, it is a very reasonable approximation to 
ignore the dependence of $E_{CM}$ on the external field.
\clearpage

%

%
\end{document}